\documentclass[graybox]{svmult}

\usepackage{natbib}
\usepackage{mathptmx}       % selects Times Roman as basic font
\usepackage{helvet}         % selects Helvetica as sans-serif font
\usepackage{courier}        % selects Courier as typewriter font
\usepackage{type1cm}        % activate if the above 3 fonts are
\usepackage{makeidx}         % allows index generation
\usepackage{graphicx}        % standard LaTeX graphics tool
\usepackage{multicol}        % used for the two-column index
\usepackage[bottom]{footmisc}% places footnotes at page bottom

\DeclareMathAlphabet{\mathcal}{OMS}{cmsy}{m}{n}

\makeindex             % used for the subject index
\newcommand{\Msun}{\ensuremath{\mathrm{M}_\odot}}

\def\kms{km\,s$^{-1}$}

\def\Ha{H$\alpha$}

\def\HeI{He\,{\sc i}}
\def\CI{C\,{\sc i}}
\def\CII{C\,{\sc ii}}
\def\OI{O\,{\sc i}}

\def\NaI{Na\,{\sc i}}
\def\MgI{Mg\,{\sc i}}
\def\MgII{Mg\,{\sc ii}}

\def\SII{S\,{\sc ii}}

\def\SiII{Si\,{\sc ii}}

\def\MgI{Mg\,{\sc i}}

\def\MgII{Mg\,{\sc ii}}
\def\CaII{Ca\,{\sc ii}}
\def\ScII{Sc\,{\sc ii}}
\def\FeI{Fe\,{\sc i}}
\def\FeII{Fe\,{\sc ii}}
\def\FeIII{Fe\,{\sc iii}}

\def\CoIII{Co\,{\sc iii}}

\def\TiII{Ti\,{\sc ii}}
\def\CrII{Cr\,{\sc ii}}
\def\SrII{Sr\,{\sc ii}}
\def\Nifs{$^{56}$Ni}
\def\Cofs{$^{56}$Co}

\def\dm15{$\Delta m_{15}(B)$}
\def\MCh{M$_\mathrm{Ch}$}
\def\lesssim{\mathrel{\hbox{\rlap{\hbox{\lower4pt\hbox{$\sim$}}}\hbox{$<$}}}}

\def\gtrsim{\mathrel{\hbox{\rlap{\hbox{\lower4pt\hbox{$\sim$}}}\hbox{$>$}}}}

\begin{document}

\title*{The Extremes of Thermonuclear Supernovae}
\label{The Extremes of Thermonuclear Supernovae}
\author{Stefan Taubenberger}
\institute{Stefan Taubenberger \at European Southern Observatory, Karl-Schwarzschild-Str. 2, 85748 Garching, Germany\\ 
                                   Max-Planck-Institut f\"ur Astrophysik, Karl-Schwarzschild-Str. 1, 85748 Garching, Germany\\ 
                                   \email{tauben@mpa-garching.mpg.de}
%\and Name of Second Author \at Name, Address of Institute \email{name@email.address}
}
\maketitle

\abstract{The majority of thermonuclear explosions in the Universe seem to proceed in a rather standardised way, as explosions of carbon-oxygen (CO) white dwarfs in binary systems, leading to `normal' Type Ia supernovae (SNe~Ia). However, over the years a number of objects have been found which deviate from normal SNe~Ia in their observational properties, and which require different and not seldom more extreme progenitor systems. While the `traditional' classes of peculiar SNe~Ia -- luminous `91T-like' and faint `91bg-like' objects -- have been known since the early 1990s, other classes of even more unusual transients have only been established 20 years later, fostered by the advent of new wide-field SN surveys such as the Palomar Transient Factory. These include the faint but slowly declining `02es-like' SNe, `Ca-rich' transients residing in the luminosity gap between classical novae and supernovae, extremely short-lived, fast-declining transients, and the very luminous so-called `super-Chandrasekhar' SNe~Ia. Not all of them are necessarily thermonuclear explosions, but there are good arguments in favour of a thermonuclear origin for most of them. The aim of this chapter is to provide an overview of the zoo of potentially thermonuclear transients, reviewing their observational characteristics and discussing possible explosion scenarios.}

\section{Introduction}
\label{Introduction}

The majority of Type Ia supernovae (SNe~Ia) form a well-defined, spectroscopically homogeneous class. Their light curves show some variation, but largely form a one-parameter family in the sense that objects with broader light curves are intrinsically more luminous, a behaviour that allows for a standardisation of their luminosity, making them valuable cosmic lighthouses. The parameter driving the light-curve width vs. peak luminosity relation is most likely the mass of radioactive \Nifs\ synthesised in the explosion, though there is growing evidence that a variation of the total ejecta mass may also be required \citep{stritzinger2006a,scalzo2014a}.

\begin{figure}[p]
\includegraphics[scale=1.0]{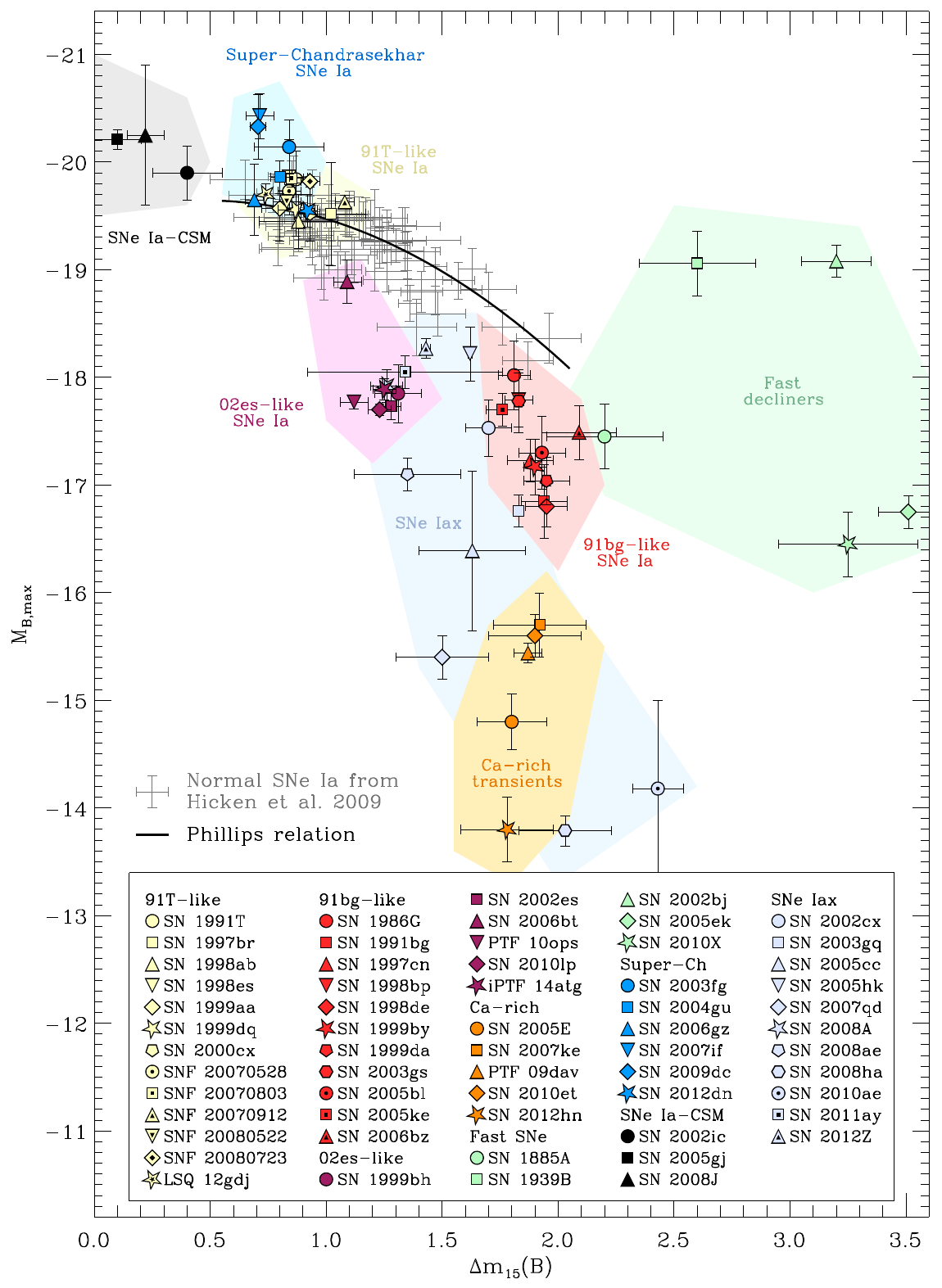}
\caption{Phase space of potentially thermonuclear transients. The absolute $B$-band magnitude at peak is plotted against the light-curve decline rate, expressed by the decline within 15\,d from peak in the $B$ band, \dm15\ \citep{phillips1993a}. The different classes of objects discussed in this chapter are highlighted by different colours. Most of them are well separated from normal SNe~Ia in this space, which shows that they are already peculiar based on light-curve properties alone. The only exception are 91T-like SNe, which overlap with the slow end of the distribution of normal SNe~Ia, and whose peculiarities are almost exclusively of spectroscopic nature. References to individual SNe are provided in the respective sections.}
\label{fig:1}
\end{figure}

However, next to these `well-behaved' SNe Ia there are also several classes of weirdos, which are spectroscopically highly peculiar or do not obey the light-curve width vs. luminosity relationship of normal SNe~Ia \citep[e.g.][]{phillips1993a,phillips1999a}. Clearly, a one-parameter description is insufficient to categorise them, and it is very likely that most of them stem from somewhat different progenitor systems and have different underlying explosion mechanisms than normal SNe~Ia. The link to the latter is that they are also believed to be thermonuclear explosions of WDs, based on the inferred nucleosynthesis or their occurrence in old stellar populations. However, some of them are clearly pushing the thermonuclear scenario to the limits. Fig.~\ref{fig:1} introduces the main classes of these more or less extreme objects and highlights their location in the light-curve width vs. luminosity plane. With the exception of `SNe~Iax' and `SNe~Ia-CSM', all other highlighted classes are discussed below. Whenever relevant, a Hubble constant $H_0 = 70$~\kms\,Mpc$^{-1}$ is assumed.

\newpage

\section{91T-like SNe: \index{91T-like SNe} spectroscopically distinct}
\label{91T-like SNe}

91T-like SNe constitute a subclass of SNe~Ia characterised by high peak luminosities and broad light curves, but especially by a peculiar pre-maximum spectroscopic evolution with \FeIII\ lines dominating the early spectra. After the formative and exceptionally well-characterised SN~1991T, a number of objects with very similar photometric and spectroscopic properties have been described in the literature. These include SN~1997br \citep{li1999a}, SNe~1998ab, 1999dq \citep{jha2006a,matheson2008a}, LSQ12gdj \citep{scalzo2014c} and several SNe from the Nearby Supernova Factory \citep{scalzo2012a,taubenberger2013a}. Since SN~1991T itself is still the best-observed of its kind, the following discussion is focused on it, before transitional 99aa-like and rare 00cx-like SNe are presented in Sections~\ref{99aa-like SNe} and \ref{00cx-like SNe}.

\begin{figure}[p]
\includegraphics[scale=1.0]{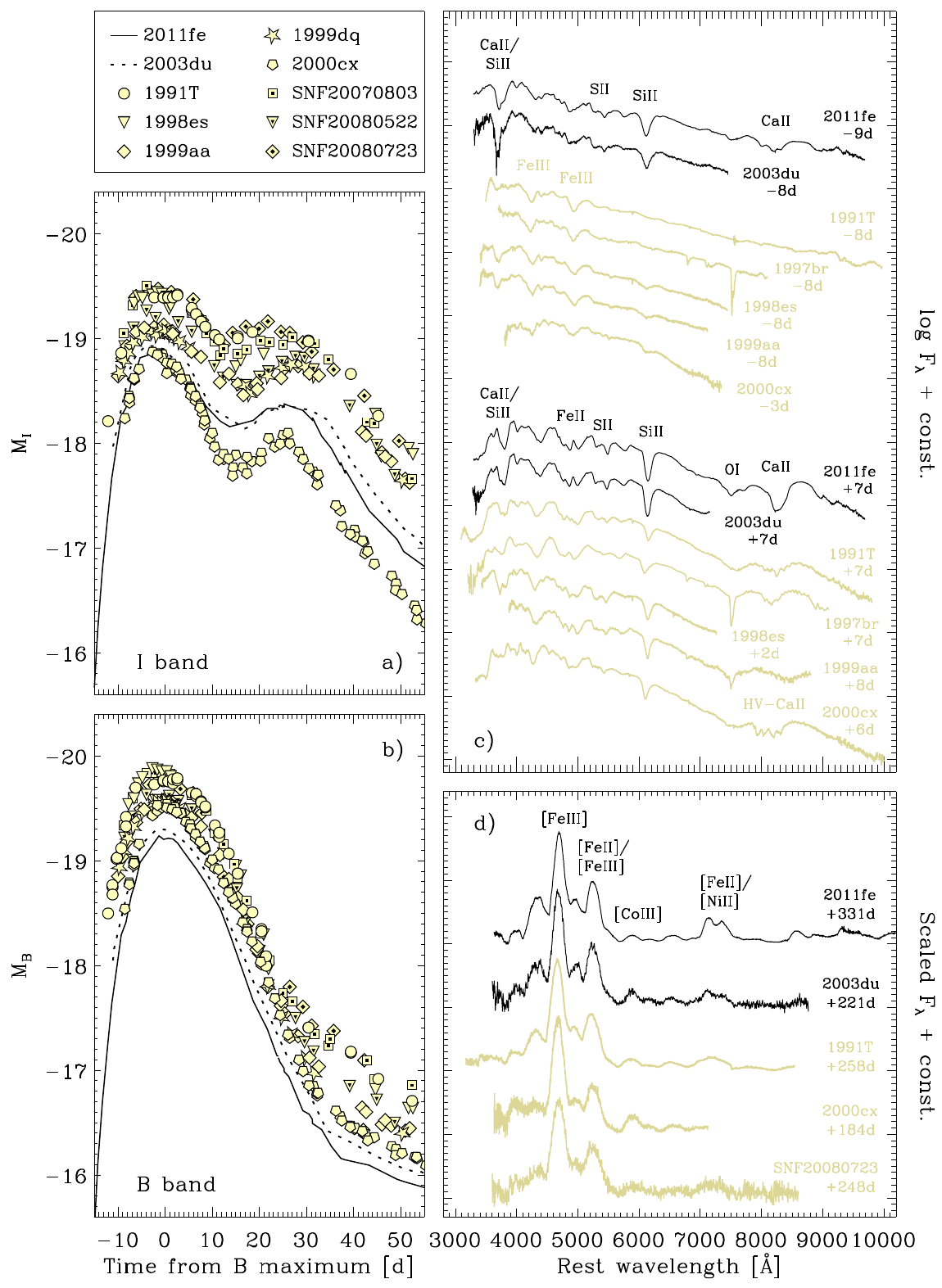}
\caption{91T-like SNe \index{91T-like SNe} in comparison to the normal SNe~Ia 2003du \citep{stanishev2007b} and 2011fe \citep{munari2013a,pereira2013a,taubenberger2015a}. a)~$I$-band light curves. b)~$B$-band light curves. c)~Photospheric spectra. d)~Nebular spectra. References for individual 91T-like SNe are provided in the main text.}
\label{fig:2}
\end{figure}

\subsection{Peak luminosity and light-curve properties}
\label{91T light curves}

Photometry of SN~1991T has been provided by \citet{filippenko1992a,phillips1992a,schmidt1994a,lira1998a,altavilla2004a}. The SN shows slowly declining light curves, with \dm15 $\sim 0.94$ \citep{phillips1999a}. The overall morphology of the light curves resembles those of normal SNe~Ia (Fig.~\ref{fig:2}a,b), and the secondary maxima in the near infrared are particularly bright \citep{hamuy1996d}. The colour evolution is rather normal, the only exception being bluer pre-maximum $U-B$ colours than in most other SNe~Ia \citep{lira1998a}.

In the original studies \citep{filippenko1992a,phillips1992a,ruiz-lapuente1992a} there was quite some controversy about the absolute peak magnitudes of SN~1991T, which originated from uncertainties in the photometric calibration, distance of the SN, and extinction within the host galaxy. All authors agreed that SN~1991T was probably more luminous than average normal SNe~Ia, but the estimated differences ranged from marginally significant 0.3\,mag \citep{phillips1992a} to $\geq 1.1$\,mag \citep{ruiz-lapuente1992a}. The corresponding \Nifs\ masses would be between 0.7 and $\geq 1.4$\,\Msun\ (assuming a \Nifs\ production of 0.5\,\Msun\ in a normal SN~Ia), the latter being incompatible with all normally considered \MCh- and sub-\MCh\ explosion models.

A better understanding of the extinction correction in SNe~Ia based on their colours \citep{lira1996a,phillips1999a}, new Cepheid-based distance estimates for the host galaxy \citep{saha2001a,altavilla2004a} and sophisticated modelling of the SN light curves and spectra \citep{jeffery1992a,mazzali1995a,sasdelli2014a} have led to some convergence in the numbers, so that nowadays peak absolute magnitudes $\sim0.4$--0.5\,mag above average SNe~Ia and \Nifs\ masses around 0.8\,\Msun\ are favoured \citep{sasdelli2014a}. This makes SN~1991T at best slightly overluminous with respect to the Phillips relation \citep{phillips1993a,phillips1999a}, but consistent within the uncertainties, as can be seen from Fig.~\ref{fig:1}. Based on a larger SN~Ia sample, \citet{blondin2012a} estimated that the class of 91T-like SNe is on average 0.2--0.3\,mag more luminous than normal SNe~Ia. 91T-like SNe are nowadays routinely included in cosmological SN~Ia samples, though some authors have warned that they might introduce a systematic bias \citep[e.g.][]{scalzo2012a}.

\subsection{Spectroscopic properties}
\label{91T spectra}

What really distinguishes SN~1991T and its siblings from other SNe~Ia is their pre-maximum spectral evolution (Fig.~\ref{fig:2}c). Early-time optical spectra of normal SNe~Ia, even of luminous and slowly declining ones, are dominated by lines of intermediate-mass elements (IMEs) such as \SiII, \SII, \CaII\ and \MgII\ \citep{branch1993a}. The \SiII\ and \CaII\ lines, in particular, are often wide and deep, and ubiquitously show high-velocity absorptions associated with them \citep{mazzali2005a}. In SN~1991T, in contrast, IME lines are essentially absent a week before maximum light, and just start to emerge around peak brightness, never reaching the same intensity as in normal SNe~Ia. Instead the early spectra are characterised by an almost featureless blue pseudo-continuum redward of $\sim$\,5500\,\AA, a high near-UV flux, and two strong absorption features near 4250 and 4950\,\AA, which have been identified as \FeIII\ multiplets $\lambda4404$ and $\lambda5129$ \citep{filippenko1992a,ruiz-lapuente1992a,jeffery1992a}. The strength of these lines requires high ionisation and a high Fe content in the outer layers of the SN ejecta, where line formation at those early epochs takes place \citep{jeffery1992a,ruiz-lapuente1992a,mazzali1995a,sasdelli2014a}. For instance, \citet{sasdelli2014a} inferred mass fractions of $\sim$3\% \Nifs\ and 1\% stable Fe in SN~1991T above 13\,000\,\kms.

Around maximum light IME lines start to emerge. \SiII\ $\lambda6355$ is clearly detected, but still weak compared to normal SNe~Ia, and the same holds for \SII\ and \CaII\ features. \SiII\ $\lambda5972$ remains virtually absent, placing SN~1991T in the shallow-silicon class following the classification scheme of \citet{branch2006a}. The expansion velocity measured from the \SiII\ $\lambda6355$ line is relatively low, and its post-peak temporal evolution very flat, so that in the classification scheme of \citet{benetti2005a} SN~1991T -- along with many normal SNe~Ia -- ends up in the low-velocity-gradient (LVG) group. Physically this may indicate that IMEs are confined to a relatively narrow region in velocity space around 10\,000\,\kms.

The further spectroscopic evolution of SN~1991T very closely resembles that of normal SNe~Ia, with spectra dominated by \FeII\ and \CaII\ lines during the weeks after maximum light. In fact, from post-maximum spectroscopy alone, 91T-like SNe can hardly be distinguished from normal SNe~Ia. Nebular spectra of SN~1991T \citep[][see Fig.~\ref{fig:2}d]{spyromilio1992a,gomez1998a,silverman2012a} reveal the usual [\FeII] and [\FeIII] emission lines. The only difference is that these lines are slightly broader than in average normal SNe~Ia, indicating that iron-group elements (IGEs) are distributed over a wide range in velocity.

\subsection{Host galaxies and relative rates}
\label{91T hosts}

91T-like SNe predominantly explode in late-type galaxies \citep{howell2001b,li2011a}, and are therefore likely associated with young stellar populations. \citet{li2011a} suggested that 91T-like SNe made up 9\% of all SNe~Ia in a volume-limited sub-sample of the Lick Observatory Supernova Search (LOSS), but given that LOSS targets mostly massive, star-forming galaxies, this estimate may be biased and the true rate of 91T-like SNe may be lower. Moreover, the 91T category of Li et al. also includes transitional, 99aa-like SNe (see Section~\ref{99aa-like SNe}). \citet{silverman2012a} estimated a much lower fraction of $\sim$2\% 91T/99aa-like SNe from the BSNIP spectroscopic SN~Ia sample, with about half of them being genuine 91T-like objects. However, the BSNIP sample is also incomplete, and the estimates may suffer from observational biases. The same is true for the sample of \citet{blondin2012a}, who found a fraction of 91T/99aa-like SNe of 5--6\% among 462 SNe~Ia.

\subsection{Explosion scenarios}
\label{91T scenarios}

Viable explosion models for SN~1991T need to produce relatively large \Nifs\ masses of $\gtrsim0.8$\,\Msun\ and reproduce the spectroscopic peculiarities described above. Spectrum-synthesis calculations have been carried out for SN~1991T using codes with different levels of physical sophistication \citep{jeffery1992a,ruiz-lapuente1992a,mazzali1995a,fisher1999a,sasdelli2014a}. All calculations agree in the need for a significant amount (a few percent) of IGEs -- \Nifs\ in particular -- in outer layers, at velocities between $\sim$10\,000 and 20\,000\,\kms, to achieve the high ionisation and reproduce the \FeIII\ absorption features observed at early phases when the interior of the SN ejecta is still optically thick. In fact, \citet{ruiz-lapuente1992a} and \citet{mazzali1995a} modelled the early spectra of SN~1991T assuming almost pure iron-peak-element compositions, but Ruiz-Lapuente et al. admitted that their modelling was energetically not self-consistent, since complete burning to IGEs would lead to higher ejecta velocities than observed.

There are basically two ways to meet the requirement to enrich the outermost layers with \Nifs: either the \Nifs\ has been produced in situ (i.e., in the outer layers of the exploding WD), or it has been produced further inside in high-density layers and mixed outward. In the early 1990s, normal SNe~Ia were still widely believed to be pure deflagrations of Chandrasekhar-mass (\MCh-) WDs. The then still relatively new delayed-detonation model \citep{khokhlov1991a}, where an initial deflagration at some point transitions into a super-sonic detonation, was considered a promising scenario to explain SN~1991T. The hope was that the detonation may burn part of the outer mantle to IGEs \citep{ruiz-lapuente1992a,mazzali1995a}. Modern three-dimensional simulations of delayed detonations with a self-consistent deflagration treatment do not confirm this behaviour. By the time the detonation sets on, the WD has had time to react to the heat released in the sub-sonic deflagration and expand. This leads to lower densities in the outer layers, and the detonation front does not burn them to IGEs, but at most to IMEs \citep[see e.g.][]{seitenzahl2013a}. 

What has instead been proposed recently \citep{fisher2015a} as a model that can lead to the inferred stratification inversion is a special variant of delayed-detonation models: a gravitationally-confined detonation \citep[GCD;][]{plewa2004a}. In this scenario the IGEs in the outermost layers are not synthesised in situ by the detonation, but are the ashes of a strongly off-centre deflagration, which rise to the WD surface in a one-sided, buoyancy-driven plume. Owing to the weak deflagration, the WD is pre-expanded only little, and an ensuing detonation can burn most of the WD interior to \Nifs, leading to very luminous events \citep{fisher2015a}. However, there is an ongoing debate whether a detonation is robustly initiated in such a scenario \citep[e.g.][]{roepke2007a}, and it remains to be tested whether the spectroscopic and photometric display of this model match the observations of 91T-like SNe.

Another idea for the origin of \Nifs\ in the outer layers suggests 91T-like SNe to be the outcome of slightly sub-\MCh\ double-detonations, with \Nifs\ produced in the initial He-shell detonation on the WD surface \citep{filippenko1992a}. Realistic simulations of this scenario \citep{fink2010a,kromer2010a} have shown that indeed the most massive exploding sub-\MCh\ WDs can reach the peak luminosity of SN~1991T, and also produce \Nifs\ in the He-shell detonation. However, synthetic observables computed for these models show little similarity with SN~1991T: strong line blanketing suppresses almost the entire flux blueward of 4000\,\AA, the velocities especially of IMEs are too high, and the light curves rise and decline too rapidly. The latter two problems can most likely only be cured by increasing the total ejecta mass, which shows that there is almost no room for a sub-\MCh\ interpretation of 91T-like SNe.

Finally, \citet{fisher1999a} promoted a super-\MCh\ scenario for SN~1991T, based on the high luminosity and the large \Nifs\ mass required. However, with the latest \Nifs\ mass estimates of $\sim$0.7--0.9\,\Msun\ \citep{scalzo2012a,sasdelli2014a}, a total mass in excess of \MCh\ may not be required. Based on a semi-analytic model with Bayesian priors, \citet{scalzo2012a} have shown that the light curve and kinetic structure of typical 91T-like SNe are compatible with a \MCh\ explosions.

\subsection{99aa-like SNe: transitional objects}
\label{99aa-like SNe}

Whenever subclasses of SNe~Ia are defined, the question arises on how distinct their properties really are. A clear-cut distinction from normal SNe~Ia could lend support to an entirely different underlying progenitor system and explosion mechanism, whereas the presence of a continuity of objects with intermediate properties might rather favour a common explosion mechanism with just some variation in certain parameters. Objects with properties intermediate between 91T-like and normal SNe~Ia have indeed been found: SN~1998es \citep{jha2006a,matheson2008a}, SN~1999aa \citep{krisciunas2000a,garavini2004a,jha2006a,matheson2008a}, and by now a number of objects akin to them. \citet{silverman2012a} have estimated the rate of 99aa-like SNe to be comparable to that of genuine 91T-like SNe.

Very early ($\sim-10$\,d) spectra of SN~1999aa resemble those of SN~1991T in being \FeIII-line dominated, and in their weakness of most IME lines. The difference to SN~1991T is that in SN~1999aa \CaII\ lines (especially H\&K) are always prominent (Fig.~\ref{fig:2}c), and that SN~1999aa transitions to a normal SN-Ia-like appearance much earlier. Already a week before maximum light, \SiII\ and \SII\ lines start to be visible, and by maximum light the SN~1999aa spectrum is very similar to those of normal SNe~Ia, with just a hint of the \FeIII\ absorptions left. This may point at an Fe-enriched, highly ionised outer layer that is less massive and extended than in SN~1991T and becomes optically thin earlier. As in 91T-like SNe, the \SiII\ $\lambda6355$ velocity measured in SN~1999aa shows little evolution from one week before to four weeks after peak brightness. Consequently, IMEs are probably confined to a narrow window in velocity space \citep{garavini2004a}.

Photometry shows that SN~1999aa is similarly luminous and slowly declining as SN~1991T, with \dm15\ measurements ranging from 0.75 \citep{krisciunas2000a} to 0.85 \citep{jha2006a}. 99aa-like SNe are routinely included in cosmological SN~Ia samples.

\subsection{00cx-like SNe: rare weirdos}
\label{00cx-like SNe}

The reason why SN~2000cx is discussed in the context of 91T-like SNe is its early-time spectroscopic appearance. Optical spectra of SN~2000cx are only available from 3\,d prior to maximum light onwards, but at that time they are still dominated by \FeIII\ lines (Fig.~\ref{fig:2}c), which are even more persistent than in SN~1991T and still detected at +15\,d \citep{li2001a}. \SiII\ and \SII\ lines are still weak in the earliest spectra, but grow significantly in strength within just a few days. It appears that SN~2000cx maintains a high ionisation and excitation in its ejecta for a longer time than most other SNe~Ia. Multi-notched high-velocity components of \CaII\ H\&K and the NIR triplet with velocities up to 20\,000\,\kms\ are prominent even one week after maximum light \citep[][see Fig.~\ref{fig:2}c]{thomas2004b,branch2004b}, again not seen in other SNe~Ia at similarly late epochs. The \SiII\ $\lambda6355$ velocity evolution shows the same flat trend as in SN~1991T, but in SN~2000cx the measured velocity is 2000--3000\,\kms\ larger \citep{li2001a}, hinting at an overall higher kinetic energy.

The photometry of SN~2000cx \citep{li2001a,candia2003a} provides no conclusive picture. The $B$-band decline is very slow [\dm15\ = 0.93], similar to that of luminous 91T-like objects. However, the rise in the $B$-band, and also the rise and decline in all other bands redward of $B$, is relatively fast (Fig.~\ref{fig:2}a,b). The secondary peak in the NIR bands is not very pronounced, and the bolometric light curve is relatively narrow, all of which is reminiscent of mildly subluminous SNe~Ia \citep{candia2003a}. Light-curve fitters fail to produce decent fits for SN~2000cx \citep{li2001a}, and absolute magnitudes estimated by them cannot be trusted. Unfortunately, the distance to NGC 524, the host galaxy of SN~2000cx, is not well constrained, and it has therefore been debated whether SN~2000cx was overluminous, normal or even subluminous \citep{li2001a,candia2003a}. The most recent calibrations actually suggest that the peak luminosity of SN~2000cx was similar to that of normal SNe~Ia \citep{silverman2013a}.

00cx-like SNe seem to be extremely rare. In fact, with its peculiar properties SN~2000cx remained unique for more than a decade, and only recently a close twin of it has been found and discussed in the literature: SN~2013bh \citep{silverman2013a}. Despite the poor statistics, it is interesting that both SN~2000cx and SN~2013bh exploded in the very outskirts of their host galaxies, at projected distances of 25 and 18 kpc, respectively, from the host centres. SN~2013bh's host is a star-forming galaxy, but at the location of the SN there is no sign of ongoing star formation, and the metallicity is likely to be low \citep{silverman2013a}. The latter is also true for SN~2000cx \citep{li2001a}, which on top of this exploded in an S0 galaxy with overall low star-formation activity. Both SNe show no signs of host-galaxy interstellar Na D absorptions in their spectra, suggesting clean circumstellar environments \citep{patat2007a,silverman2013a}. All this might indicate that 00cx-like SNe stem from relatively old progenitors, in sharp contrast to 91T- and 99aa-like SNe.

\newpage

\section{91bg-like SNe: \index{91bg-like SNe} cool and subluminous}
\label{91bg-like SNe}

In 1991 the paradigm of SN~Ia homogeneity changed drastically, since not only the luminous and spectroscopically unusual SN~1991T was discovered (see Section~\ref{91T-like SNe}), but also the  subluminous, fast-declining SN~1991bg \citep{filippenko1992b,leibundgut1993a,turatto1996a}. SN~1991bg became the prototype of a whole class of objects, some of which have properties nearly identical to SN~1991bg, while others seem to bridge the gap to normal SNe~Ia, and again others deviate in certain respects and are sometimes even more extreme than SN~1991bg itself. In the following we will first focus on SN~1991bg and its close twins, which include SNe~1997cn \citep{turatto1998a,jha2006a}, 1998de \citep{modjaz2001a,jha2006a,matheson2008a}, 1999by \citep{howell2001a,vinko2001a,hoeflich2002b,garnavich2004a,silverman2012a}, 2005bl \citep{taubenberger2008a}, 2005ke \citep{hicken2009b,contreras2010a,blondin2012a,silverman2012a,patat2012a}, 2006bz \citep{hicken2009b}, and several other examples contained in the SN~Ia samples of \citet{hicken2009b}, \citet{contreras2010a}, \citet{ganeshalingam2010a}, \citet{stritzinger2011a}, \citet{blondin2012a} and \citet{silverman2012a}. In Sections~\ref{86G-like SNe} and \ref{PTF09dav} we will then discuss transitional 86G-like objects and the very subluminous and peculiar transient PTF\,09dav, respectively.

\begin{figure}[p]
\includegraphics[scale=1.0]{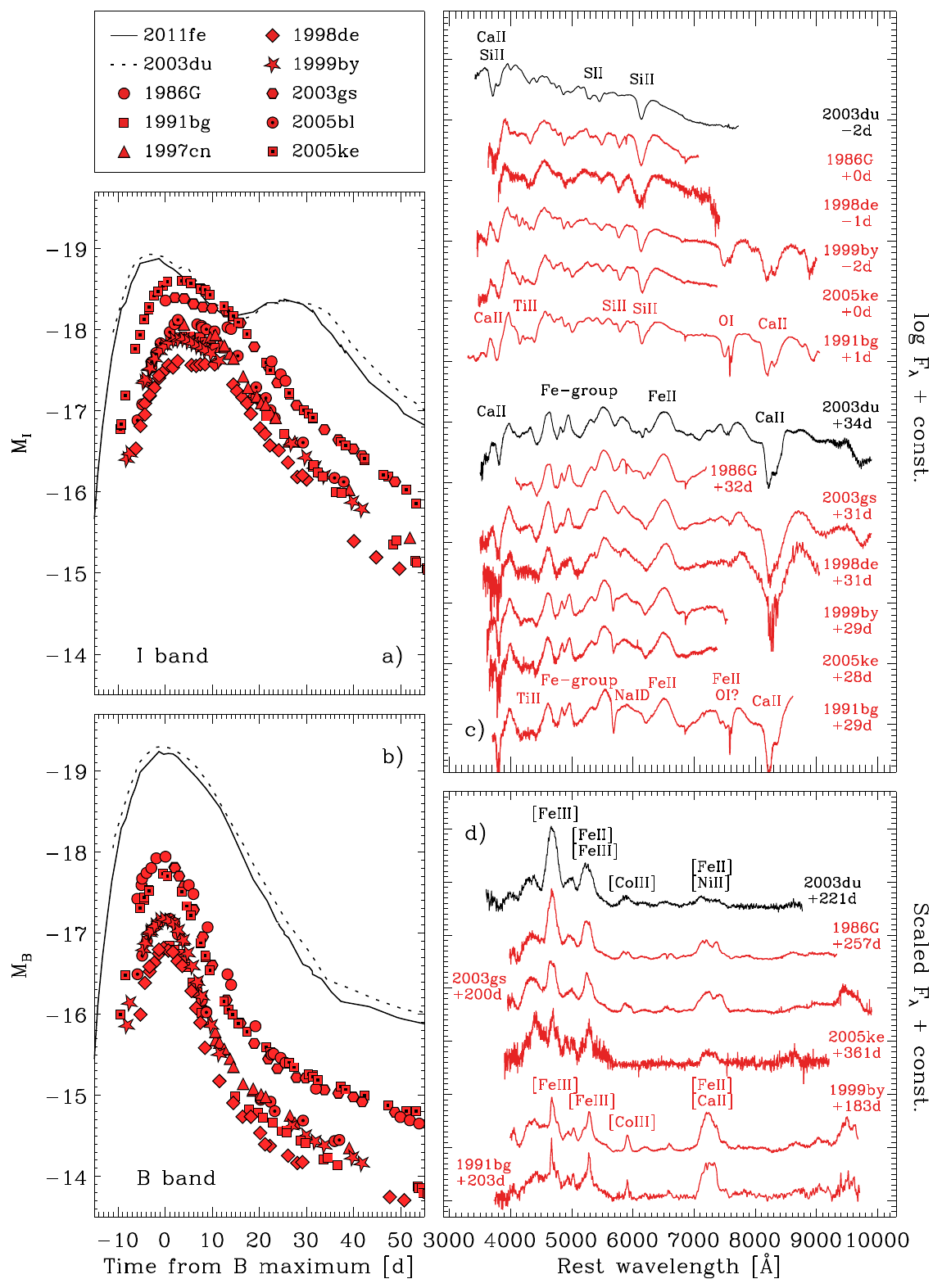}
\caption{91bg-like SNe \index{91bg-like SNe} in comparison to the normal SNe~Ia 2003du \citep{stanishev2007b} and 2011fe \citep{munari2013a,pereira2013a}. a)~$I$-band light curves. b)~$B$-band light curves. c)~Photospheric spectra. d)~Nebular spectra. References for individual 91bg-like SNe are provided in the main text.}
\label{fig:3}
\end{figure}

\subsection{Luminosity and colour evolution}
\label{91bg light curves}

The most obvious property of 91bg-like SNe is their significantly lower luminosity compared to normal SNe~Ia. At peak they are 1.5--2.5\,mag fainter in optical bands, reaching peak absolute magnitudes between $-16.7$ and $-17.7$ in $B$ \citep[][see Fig.~\ref{fig:1}]{taubenberger2008a,sullivan2011a}. With rather red maximum-light colours, $(B-V)_\mathrm{max} \sim0.5$--0.6\,mag \citep{taubenberger2008a}, they are more reminiscent of stripped-envelope core-collapse SNe than normal SNe~Ia, and while normal SNe~Ia are reddest $\sim30$\,d after maximum light, 91bg-like SNe reach this point earlier, at $\sim+15$\,d \citep{burns2014a}. Only a month or more after maximum light, during the \FeII-dominated phase, the colours of all SNe~Ia converge to a common value \citep{lira1996a}.

At longer wavelengths, the luminosity differences are less pronounced. \citet{wood-vasey2008a} found indistinguishable $JHK$ peak absolute magnitudes for normal and subluminous SNe. \citet{kattner2012a} instead suggested that there is a weak but significant trend of NIR peak absolute magnitudes with decline rate, and that 91bg-like SNe are fainter than normal SNe~Ia also in $J$ and $H$, but only by about 0.5\,mag. Finally, \citet{krisciunas2009a} concluded that there are two groups of fast-declining SNe~Ia: transitional objects (Section~\ref{86G-like SNe}) have NIR peak magnitudes indistinguishable from those of normal SNe~Ia, whereas genuine 91bg-like objects are fainter than normal SNe~Ia by 0.5--0.7\,mag, confirming an earlier result for SN~1999by by \citet{garnavich2004a}. All these studies, however, are limited by the small number of 91bg-like SNe in the investigated samples.

The light curves of 91bg-like SNe show a fast decline from peak, reflected in a light-curve decline parameter \dm15\ of 1.8--2.1 [\dm15\,$\lesssim 1.7$ in normal SNe~Ia]. From pre-maximum light curves rise times of 13 to 15\,d have been estimated \citep{taubenberger2008a}, a few days shorter than those of normal SNe~Ia (Fig.~\ref{fig:3}a,b). However, a simple time stretch is insufficient to map the light curves of 91bg-like SNe onto those of normal SNe~Ia. First and foremost, 91bg-like SNe lack the secondary maxima in the NIR bands that are characteristic of normal SNe~Ia (Fig.~\ref{fig:3}a). Instead, they show single peaks delayed by a few days with respect to that in $B$ \citep{garnavich2004a,gonzalez-gaitan2014a,friedman2015a}. The first peak in normal SNe~Ia, in contrast, precedes those in bluer bands. \citet{kasen2006b} and \citet{blondin2015a} suggested the occurrence of the secondary NIR peak in SNe~Ia to be related to a recombination wave propagating through chemically stratified ejecta. The recombination of \FeIII\ to \FeII\ would happen earlier in less luminous and cooler SNe, and in 91bg-like SNe this would cause the first and second NIR peaks to merge to form a single, slightly delayed maximum.

Starting at \dm15\,$\sim1.7$, many quantities of SNe~Ia such as colours and spectroscopic appearance (see below) show a steeper dependence on \dm15\ than at lower decline rates. This also holds for the peak luminosity, making 91bg-like SNe~Ia fainter than predicted by the Phillips relation for the given \dm15\ \citep[][see Fig.~\ref{fig:1}]{garnavich2004a,taubenberger2008a}. This deviation from the width\,--\,luminosity relation as calibrated with normal SNe~Ia disqualifies 91bg-like SNe as useful cosmic distance indicators. However, since at higher redshift they are disfavoured by Malmquist bias anyway, they are unlikely to compromise SN~Ia cosmology.

The inferred quasi-bolometric peak luminosities of 91bg-like SNe are at least a factor 2--3 below those of normal SNe~Ia, and range from log($L$)\,=\,42.1 to 42.4 \citep[e.g.][]{taubenberger2008a}. The bolometric peak can be used to estimate the mass of \Nifs\ synthesised in an explosion. For 91bg-like SNe, \Nifs\ masses between $\sim$0.05 and $\sim$0.10\,\Msun\ \citep[e.g.][]{mazzali1997a,sullivan2011a} have been derived in this way. The late-time tail of bolometric light curves is sensitive to the \Nifs\ mass as well, but also to the degree of $\gamma$-ray and positron trapping.  Normal SN~Ia bolometric light curves can be reproduced assuming complete positron trapping, but increasing $\gamma$-ray losses owing to the expansion of the ejecta \citep[e.g.][]{kerzendorf2014a}. There are indications that this scheme breaks down for 91bg-like SNe \citep{mazzali1997a}. Either there is increasing flux redistribution into bands not accounted for in the quasi-bolometric light curve, or positron trapping is incomplete, which would have important consequences for the magnetic-field configuration in the ejecta \citep{ruiz-lapuente1998a}, and might hint at an explosion mechanism different from normal SNe~Ia.

\subsection{Spectroscopic characteristics}
\label{91bg spectra}

The spectroscopic evolution of 91bg-like SNe largely resembles that of normal SNe~Ia. Pre-maximum optical spectra are dominated mostly by intermediate-mass elements, but soon after maximum light they transition to being shaped by Fe and Co lines as the line-forming region recedes into the IGE-rich core of the ejecta (Fig.~\ref{fig:3}c). This transition happens earlier in 91bg-like SNe than in normal SNe~Ia, but otherwise the similarities in the evolution indicate a comparable degree of chemical stratification in the ejecta. In particular, strong outward-mixing of IGEs is disfavoured by the Fe-poor early spectra.

However, looking in detail, a number of spectroscopic peculiarities in 91bg-like SNe become evident \citep{filippenko1992b,leibundgut1993a,mazzali1997a}. In early-time spectra \OI\ $\lambda7774$ and \TiII\ lines between 4000 and 4400\,\AA\ are unusually prominent, and the Ti lines increase in strength until few days past peak brightness (Fig.~\ref{fig:3}c). \CI\ lines have clearly been detected in NIR spectra \citep{hoeflich2002b,hsiao2015a}, \CII\ lines more tentatively in optical spectra as late as a few days before maximum \citep{taubenberger2008a,blondin2012a}. \SiII\ lines are pronounced, and \SiII\ $\lambda5972$ is particularly strong \citep{nugent1995a,hachinger2008a}. The characteristic \SII\ `W' feature, in contrast, appears less marked than in normal SNe~Ia. Some of the observed peculiarities can be attributed to the comparatively low temperature and ionisation state of the ejecta \citep{mazzali1997a,hachinger2008a,hachinger2009a}, which leads to a higher abundance of neutral and singly ionised species and disfavours lines arising from highly excited states. However, it is unlikely that the temperature alone can explain all peculiarities. A truly higher abundance of oxygen is probably required \citep{hachinger2009a}, but also not unexpected in a subluminous SN with a low \Nifs\ mass and overall low burning efficiency. Whether or not an enhanced Ti abundance is needed is more controversial. It would also be much harder to explain from a nucleosynthesis perspective.

The ejecta velocities measured in 91bg-like SNe are on average lower, but not dramatically different from those in normal SNe~Ia. Before maximum light the velocities estimated from \SiII\ $\lambda6355$ are on the low side (but still within the range) of velocities observed in normal SNe~Ia. After maximum light, however, when in low-velocity normal SNe~Ia the \SiII\ velocities settle at 9000--10\,000\,\kms, those in 91bg-like SNe keep decreasing, reaching typical values around 7000\,\kms\ \citep{taubenberger2008a} by the time the \SiII\ $\lambda6355$ line disappears because of blending with emerging \FeII\ emission. Up to now, no high-velocity features (HVFs; \citealt{mazzali2005a}) have ever been detected in 91bg-like SNe. Since the origin of HVFs in SNe~Ia is still obscure, it is also unclear whether the absence of HVFs in 91bg-like SNe is related to their explosion mechanism or to their circumstellar environments.

Nebular ($\gtrsim100$\,d) spectra of SNe~1991bg and 1999by are again rather unusual (Fig.~\ref{fig:3}d). While normal SNe~Ia are characterised by a mix of broad [\FeII], [\FeIII] and [\CoIII] emission lines at those phases, broad lines of doubly ionised species are absent in SNe~1991bg and 1999by. Their nebular spectra are dominated by broad [\FeII] and [\CaII] $\lambda\lambda7291,7324$ emission lines, indicative of a comparatively low ionisation throughout most of the ejecta. However, \textit{narrow} [\FeIII] and [\CoIII] emission lines (fwhm\,=\,1000--2000\,\kms) are superimposed on the underlying broad [\FeII] spectrum, and their relative strength increases with time \citep{turatto1996a,mazzali1997a}. Their width suggests that they originate exclusively from the innermost ejecta. \citet{mazzali2012a} argued that the required high ionisation state in this innermost region could be explained by a low central density, as it is found in violent-merger explosion models \citep{pakmor2010a}.

Based on their fast light-curve decline and their strong \SiII\ $\lambda5972$ line, 91bg-like SNe belong to the class of CL (`cool') SNe in the Branch et al. classification scheme \citep{branch2006a}, and to the class of `FAINT' SNe following \citet{benetti2005a}. In fact, 91bg-like SNe form the cores of these groups, which encompass also transitional objects and some weirdos, and whose boundaries to other SN~Ia classes are not sharp.

\subsection{Polarimetry}
\label{91bg polarimetry}

Spectropolarimetric observations are available for SNe~1999by \citep{howell2001a} and 2005ke \citep{patat2012a}, and reveal interesting differences to the polarisation signal observed in normal SNe~Ia. The latter typically show very low continuum polarisation ($\leq$0.2\%), indicating that the ejecta show no global deviation from a spherical shape. At the same time, strong line polarisation (up to 1\%) is observed in \SiII\ and \CaII\ lines in early-time spectra, pointing at significant compositional inhomogeneities in the outer layers \citep{wang2007a}. HFVs, in particular, are always strongly polarised. From this point of view it may not be surprising that in SNe~1999by and 2005ke line polarisation is among the weakest in SNe~Ia \citep[$\sim$0.4\%;][]{howell2001a,wang2007a,patat2012a}: 91bg-like SNe do not show HVFs. What is really remarkable, however, is the relatively high continuum polarisation in SNe~1999by and 2005ke, which reaches and even exceeds 0.5\% in the line-free region around 7000\,\AA\ \citep{howell2001a,patat2012a}. According to Patat et al., this could be explained by an oblate geometry with a global asymmetry at the 15\% level. Suitable explosion models for 91bg-like SNe must be capable of producing that degree of asphericity.

\subsection{Host galaxies and relative rates}
\label{91bg hosts}

In stark contrast to 91T-like SNe which explode almost exclusively in star-forming late-type galaxies (see Section~\ref{91T hosts}), 91bg-like SNe are mostly found in massive elliptical or S0 galaxies with a star-formation rate below a few times $10^{-9}$\,\Msun\,yr$^{-1}$ \citep{howell2001b,neill2009a,gonzalez-gaitan2011a}. Only few 91bg-like SNe have been found in spiral galaxies, most prominently SN~1999by, which exploded in the Sb galaxy NGC\,2841. However, despite its morphological classification the spectral features of NGC\,2841 are those of an elliptical galaxy \citep{gallagher2005a}. 
 
Concerning the relative rate of 91bg-like SNe, discordant numbers have been quoted in the literature. Based on a volume-limited subsample of SNe discovered by LOSS, \citet{li2011a} estimated that 91bg-like SNe constitute 15\% of all SNe~Ia. \citet{silverman2012a} obtained a much lower fraction of 91bg-like SNe of $\sim$6\% of all SNe~Ia on the basis of the Berkeley Supernova Ia Program (BSNIP) spectroscopic sample. Using only light-curve information, \citet{ganeshalingam2010a} derived a fraction of $\sim$11\% for a largely congruent sample of SNe. Finally, \citet{gonzalez-gaitan2011a} estimated that 7--9\% of all local SNe~Ia are 91bg-like. All these studies have different systematics, and all have a strong host-galaxy bias since they are based on samples discovered by targeted SN surveys. However, the biggest uncertainty in the rate comes from the demarcation of 91bg-like SNe, and the question whether transitional, 86G-like SNe (Section~\ref{86G-like SNe}) are counted separately or not. Including transitional objects, the fraction of subluminous SNe~Ia estimated by \citet{gonzalez-gaitan2011a} triples, from 7--9\% to 22\%.
In magnitude-limited samples, 91bg-like SNe are strongly suppressed by Malmquist bias. Accordingly, \citet{oestman2011a} and \citet{foley2009b} did not find any 91bg-like SNe~Ia in their higher-redshift samples with mean $z$ of 0.17 and 0.35, respectively, and while \citet{gonzalez-gaitan2011a} did find several subluminous SNe~Ia in the SNLS data set at $0.1 \lesssim z \lesssim 1.0$, with one exception they were all transitional objects.

\subsection{Explosion scenarios}
\label{91bg scenarios}

Promising explosion models for 91bg-like objects need to explain the main characteristics of these SNe, including the low \Nifs\ masses, the rapidly declining light curves, the relatively high continuum polarisation and the particular chemical abundance structure. Spectral modelling carried out by \citet{hachinger2009a} revealed that the ejecta of 91bg-like SNe are chemically stratified, dominated by oxygen above $\sim$8000\,\kms\ and by IMEs below. In particular, strong outward mixing of IGEs is not compatible with the spectroscopic display of 91bg-like SNe. This result basically rules out pure deflagrations \citep[e.g.][and references therein]{fink2014a} as a possible explosion mechanism for 91bg-like objects, although the required low \Nifs\ masses could be accomplished within this scenario. Delayed detonations \citep[e.g.][and references therein]{seitenzahl2013a}, on the other hand, retain a high level of chemical stratification, but fail to produce \Nifs\ masses below $\sim$0.2--0.3\,\Msun, not low enough for 91bg-like SNe. Taken together, 91bg-like objects are unlikely to originate from thermonuclear explosions in \MCh\ WDs.

\citet{pakmor2010a} presented a violent merger of two 0.9\,\Msun\ WDs, which produced about the right \Nifs\ mass and spectroscopically provided a satisfactory match with 91bg-like objects. However, because of the large total ejecta mass of 1.8\,\Msun\ and the accordingly large optical depths for $\gamma$-rays and optical photons, the light curve of that model was very broad and inconsistent with those of rapidly declining 91bg-like events. In fact, using an analytic light-curve model, \citet{stritzinger2006a} and \citet{scalzo2014a} claimed significantly sub-\MCh\ progenitors for fast decliners.

Sub-\MCh\ explosions are typically realised as double detonations, where a first detonation in an accreted He shell on top of the WD triggers a second detonation in the CO core via converging shocks \citep[][and references therein]{fink2010a}. If the exploding WD is sufficiently lightweight, only small amounts of \Nifs\ are produced. The main problem in the classical double-detonation models is the effect of the ashes of the He detonation: they are rich in IGEs and generate a lot of opacity in the outer layers, which is usually detrimental to the spectrum \citep{kromer2010a}. This effect might be mitigated or avoided by keeping the He-shell mass at a minimum, and by igniting the detonation dynamically through accretion-stream instabilities \citep{guillochon2010a} or in the course of a violent merger \citep{pakmor2013a}. As a matter of fact, even the very-low-mass He shells natively present on top of virtually all CO WDs might be sufficient for these scenarios to work. However, the capability of these models to reproduce 91bg-like SNe  still remains to be demonstrated.

\subsection{86G-like SNe: half way between normal and 91bg-like SNe}
\label{86G-like SNe}

Five years before SN~1991bg, the first peculiar SN~Ia was discovered: SN~1986G in Centaurus A exhibited a low luminosity and unusually red colours \citep{phillips1987a}. However, since the SN was embedded in a dust lane, and showed prominent interstellar \NaI\,D absorption lines in its spectra, it was not totally clear to which extent the observed peculiarities were intrinsic, or just caused by extinction and reddening along the line of sight.  With a better handle on the extinction towards SN~1986G \citep{hough1987a,phillips1999a}, it has become clear that the SN is indeed intrinsically unusual, and seems to be a transitional object between normal and 91bg-like SNe. Several other 86G-like SNe have subsequently been studied, including SN~1998bp \citep{jha2006a,matheson2008a}, SN~2003gs \citep{krisciunas2009a,silverman2012a} and iPTF\,13ebh \citep{hsiao2015a}. They appear to be at least as frequent as the more extreme 91bg-like objects \citep{gonzalez-gaitan2011a}.

The light curves of 86G-like SNe decline rapidly [with a typical \dm15\ around 1.8, see Fig.~\ref{fig:1}], and peak at magnitudes intermediate between those of normal and 91bg-like objects ($M_{B,\mathrm{max}} \sim -18$, translating into \Nifs\ masses of about 0.2\,\Msun). In NIR bands, 86G-like SNe do show double-peaked light curves, with the first peak preceding those in bluer bands, in analogy to normal SNe~Ia. However, the secondary maxima are very weak, and occur soon after the first peaks. The $JHK$ peak magnitudes are brighter than in 91bg-like SNe, and align with those of normal SNe~Ia, which have been proposed to be standard candles in the NIR \citep{wood-vasey2008a,krisciunas2009a,kattner2012a}.

Spectroscopically, 86G-like SNe show again properties intermediate between normal and 91bg-like events. The \SiII\ $\lambda5972$ line is as strong as in 91bg-like objects, but \TiII\ lines -- while present -- are considerably weaker, indicating an intermediate ionisation state (Fig.~\ref{fig:3}c). At nebular phases \citep{cristiani1992a,silverman2012a}, 86G-like objects show a high ionisation in a sufficiently large volume of the ejecta to produce broad [\FeIII] emission (Fig.~\ref{fig:3}d), in contrast to the narrow [\FeIII] emission seen in 91bg-like objects which arises exclusively from the innermost regions \citep{mazzali2012a}.

With their intermediate properties, 86G-like objects play a critical role for understanding whether subluminous SNe~Ia are just the extreme tail of a single SN~Ia population, or a distinct class of objects with different progenitors or explosion mechanisms.

\subsection{PTF\,09dav: simply the most subluminous 91bg-like event?}
\label{PTF09dav}

PTF\,09dav \citep{sullivan2011a} may or may not constitute the most extreme 91bg-like SN to date by quite a margin. Even though the eventual classification remains controversial, there are reasons to consider PTF\,09dav a 91bg-like event. First of all, the light-curve morphology with a single peak in the $I$ band and a rapid decline in all bands is very similar (Fig.~\ref{fig:5}a). \citet{sullivan2011a} estimated a \dm15\ of $1.87\pm0.06$, well within the range observed in 91bg-like objects. The colours at maximum light are also the same as in 91bg-like SNe, with $(B-V)_\mathrm{max}=0.56$\,mag, and the main spectroscopic features in form of prominent \TiII, \SiII\ and \OI\ lines are identical (Fig.~\ref{fig:5}c).

What makes PTF\,09dav stand out from the crowd of 91bg-like objects is its extremely low luminosity. With a peak absolute $B$-band magnitude of about $-15.5$, it is $\sim1.5$\,mag fainter than other 91bg-like SNe, and $\sim3.5$\,mag fainter than normal SNe~Ia (Fig.~\ref{fig:1}), leading to a \Nifs\ mass estimate of merely 0.019\,\Msun\ \citep{sullivan2011a}.

The low luminosity goes hand in hand with a number of peculiarities in early-time spectra, which \citet{sullivan2011a} have analysed in detail. To start with, the derived expansion velocities are low, even by the standards of subluminous SNe~Ia, reaching only $\sim$6000\,\kms\ a few days after maximum light. \SiII\ $\lambda5972$ is surprisingly weak, in spite of the apparently low temperature and ionisation of the ejecta. This might be explained by an overall low Si abundance, which would be in line with the complete absence of \SII\ lines in PTF\,09dav. Instead, spectral modelling suggests the presence of a number of species that require low temperatures: \TiII, \MgI, \NaI\ and \ScII. The \ScII\ detection, in particular, appears convincing, and is unprecedented in thermonuclear SNe, where the abundance of Sc is expected to be orders of magnitude below those of e.g. Ca, Ti and Cr \citep[e.g.][]{seitenzahl2013a}. The tentative identification of \SrII\ lines -- if correct -- would be even more difficult to explain within a thermonuclear explosion scenario, as Sr is a classical s-process element. Interestingly, there is almost no Fe required to obtain a good fit. This is especially true at more advanced phases ($\sim2$\,weeks after maximum), when normal 91bg-like SNe start to be dominated by \FeII\ emission. Finally, a nebular spectrum of PTF\,09dav taken three months after maximum \citep{kasliwal2012a} shows little else than strong, broad [\CaII] $\lambda\lambda7291,7324$ emission and weak, narrow \Ha\ (Fig.~\ref{fig:5}d). Fe emission lines, which dominate the nebular spectra of all other SNe~Ia including 91bg-like objects, are entirely absent. 

It may be doubted that all the observed spectroscopic peculiarities can be solely explained by a lower ejecta temperature (though this may indeed be an important factor). Instead, it appears likely that the nucleosynthesis in PTF\,09dav proceeded differently than in other SNe~Ia, favouring elements between Ca and Cr at the expense of both classical IMEs such as Si and S, and classical IGEs such as Fe, Co and Ni. The consequence of this would be that the similarities of PTF\,09dav with 91bg-like SNe at early phases are mostly coincidental, that it likely had a different progenitor and explosion mechanism, and should not be regarded as a member of this group. We will come back to PTF\,09dav in the context of the so-called Ca-rich gap transients \citep{kasliwal2012a} in Section~\ref{Ca-rich}, where we compare its properties with those of other low-luminosity SNe with predominant [\CaII] emission at late phases.

\newpage

\section{02es-like SNe: \index{02es-like SNe} subluminous but slowly declining}
\label{02es-like}

With `02es-like SNe' we refer to a group of objects which are not really extreme in any one photometric or spectroscopic property, but populate a region in SN~Ia phase space that was long believed to be devoid of objects, and which lies far off the Phillips relation (Fig.~\ref{fig:1}).

The first observations of an 02es-like SN date back to the end of the 1990s, but more than 10\,yr should pass before the first data were published. 02es-like SNe are hence the most recent addition to the zoo of thermonuclear transients. Actually, the group is so young that not even the term `02es-like SNe' is well established in the literature, let alone a sharp membership definition. What is used here is a somewhat wider reading of the term compared to \citet{white2015a}, including also SNe that do not exhibit the same very low ejecta velocities as SN~2002es. With this definition the class encompasses SN~2002es itself and SN~1999bh \citep{li2001b,ganeshalingam2012a}, SN~2010lp (Kromer et al. 2013, Taubenberger et al. 2013b), PTF\,10ops \citep{maguire2011a}, iPTF\,14atg \citep{cao2015a,kromer2016a}, PTF\,10ujn and PTF\,10acdh \citep{white2015a}, and arguably SN~2006bt \citep{hicken2009b,ganeshalingam2010a,foley2010a}. \citet{white2015a} have  shown that 02es-like SNe are clearly distinct from SNe~Iax in a number of key properties.

\begin{figure}[p]
\includegraphics[scale=1.0]{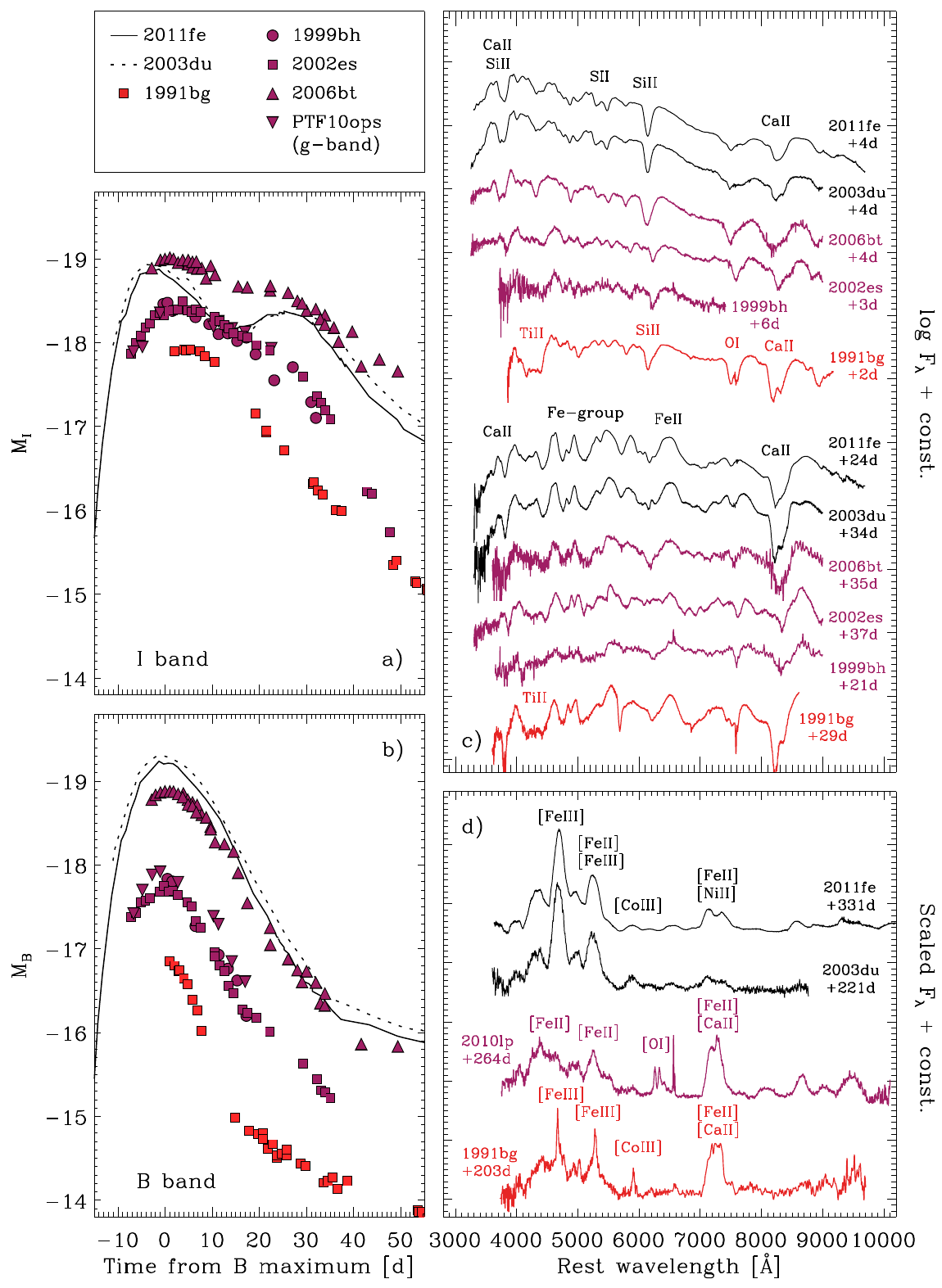}
\caption{02es-like SNe \index{02es-like SNe} in comparison to the normal SNe~Ia 2003du \citep{stanishev2007b} and 2011fe \citep{munari2013a,pereira2013a,taubenberger2015a}, and the subluminous SN~Ia 1991bg \citep[][see also Section~\ref{91bg-like SNe}]{filippenko1992b,leibundgut1993a,turatto1996a}. a)~$I$-band light curves. b)~$B$-band light curves ($g$-band for PTF\,10ops). c)~Photospheric spectra. d)~Nebular spectra. References for individual 02es-like SNe are provided in the main text.}
\label{fig:4}
\end{figure}

\subsection{General evolution and diversity}
\label{02es-like properties}

The description of the observational characteristics of 02es-like SNe can be strongly abbreviated by referring to Section~\ref{91bg-like SNe}, where the spectroscopic and photometric properties of 91bg- and 86G-like SNe are presented. This is because 02es-like SNe behave very similarly to 91bg\,/\,86G-like SNe in most respects, featuring cool ejecta with low ionisation and spectra with strong \SiII\ (especially $\lambda5972$), \OI\ and \TiII\ lines at early phases (Fig.~\ref{fig:4}c). The light-curve morphology with a single peak in the $I$ band also resembles that of traditional subluminous SNe (Fig.~\ref{fig:4}a), and so do the relatively red colours [$(B-V)_\mathrm{max}=0.2\,...\,0.5$]. The absolute magnitudes are faint, with $M_{B,\mathrm{max}}$ ranging from $-17.6$ to $-18.1$ for most of the objects, more similar to SN~1986G than to the fainter genuine 91bg-like SNe, but still significantly subluminous compared to normal SNe~Ia. Through Arnett's rule \citep{arnett1982a} or more sophisticated light-curve modelling, \Nifs\ masses around 0.15--0.20\,\Msun\ have been estimated \citep{maguire2011a,ganeshalingam2012a,kromer2013b}. Only SN~2006bt is an exception, peaking at $M_B\approx-19$ and thus being almost as luminous as normal SNe~Ia with a similar \dm15\ of about 1.1 \citep[][see Figs.~\ref{fig:1} and \ref{fig:4}b]{foley2010a}.

What really distinguishes 02es-like objects from 91bg- and 86G-like SNe are their time scales. Their light-curve peaks are much wider, with longer rise times of 19--20\,d \citep{maguire2011a,cao2015a} and slower decline. In fact, with \dm15\ in the range 1.1 to 1.3, 02es-like SNe in this respect closely resemble normal SNe~Ia, and are very different from 91bg-like objects with \dm15\ between 1.8 and 2.0. The overall stretched light-curve peak is paralleled by a slower evolution of the SN colours, and by a later and more gentle transition from IME- to \FeII-dominated spectra.

A point in which the class of 02es-like SNe shows significant diversity are the ejecta velocities. Some objects like SN~2006bt or PTF\,10ops show \SiII\ $\lambda6355$ velocities at peak of $\sim10\,000$\,\kms\ \citep{foley2010a,maguire2011a}, while SN~2002es and its closest siblings (SN~1999bh and PTF\,10acdh) show \SiII\ velocities between 6000 and 7000\,\kms\ \citep{ganeshalingam2012a,white2015a}. With $\sim8000$\,\kms, iPTF\,14atg takes an intermediate position \citep{cao2015a}.

\subsection{Early UV spike in iPTF\,14atg}
\label{iPTF14atg UV}

Among all 02es-like SNe, iPTF\,14atg is the only one which was discovered very soon after explosion, and which has an excellent coverage of the early light curve by ground-based and \textit{swift} UV photometry. In the earliest photometric data points, up to $\sim5$\,d after the estimated instant of explosion, it shows a spike in the \textit{swift} UV bands, which is also tentatively visible in the $u$ and $b$ bands \citep{cao2015a}. No excess is visible in redder bands, in particular in the $r$ band, even at an extremely early phase $\sim1$\,d after the explosion. A spectrum of iPTF\,14atg taken $\sim3$\,d after explosion \citep{cao2015a}, during the UV spike, shows a blue continuum with only relatively weak, broad features which do not seem to match any of the lines seen later on. Without similarly early observations for other 02es-like SNe, it is unclear whether or not this behaviour is generic for this class of objects.

\subsection{Nebular spectroscopy of SN~2010lp}
\label{10lp nebular}

A nebular spectrum of SN~2010lp (264\,d after maximum; \citealt{taubenberger2013b}) in most respects closely resembles spectra of 91bg-like SNe taken at similar phases (see Section~\ref{91bg spectra}), being dominated by broad [\FeII] and [\CaII] emission (Fig.~\ref{fig:4}d). No broad [\FeIII] emission is visible, and the curious \textit{narrow} ($\sim2000$\,\kms FWHM) [\FeIII] and [\CoIII] emission lines, which were detected in SNe~1991bg and 1999by and gave rise to speculations on a low central matter density \citep{mazzali2012a}, are totally absent in SN~2010lp. Instead, [\OI] $\lambda\lambda6300,6364$ emission is observed there \citep{taubenberger2013b}, which is even more curious for a thermonuclear SN, as these lines are usually considered hallmark features of core-collapse explosions. The [\OI] lines in SN~2010lp are similarly narrow as the [\FeIII] lines in SN~1991bg and double-peaked, which suggests the emission to arise from a complexly shaped oxygen-rich region in the central parts of the ejecta. Again, it remains to be seen how generic this behaviour is among 02es-like SNe, although a low-S/N nebular spectrum of iPTF\,14atg \citep{kromer2016a} seems to show the same features as that of SN~2010lp.

\subsection{Rapid fading of SN~2002es after one month}
\label{02es fading}

About 30\,d after maximum, just before the expected settling on a radioactive tail, all optical light curves of SN~2002es start to fade rapidly \citep[][see Fig.~\ref{fig:4}a,b]{ganeshalingam2012a}. The enhanced decline lasts for at least 40\,d (when the photometric follow-up ends). Between $+30$ and $+70$\,d, SN~2002es fades by 3.5--4.5\,mag in $B$, $V$ and $R$, about three times as much as a normal SN~Ia in the same period. Qualitatively, the behaviour is similar to the decline observed in some `super-Chandrasekhar' SNe~Ia at somewhat later epochs (see Section~\ref{super-Ch}), which there has been attributed to dust formation \citep{taubenberger2013a}. \citet{ganeshalingam2012a} discussed several possible explanations, including dust formation, an IR catastrophe \citep{axelrod1980a}, $\gamma$-ray escape and a light-curve peak powered by shock breakout rather than radioactivity. $\gamma$-ray escape would require none of $\gamma$-rays and only 30\% of the positrons to be trapped at $+70$\,d, which is nearly impossible if the trapping was still mostly complete at maximum light. Dust formation at such an early epoch would be unprecedented, and is disfavoured by the fact that the SN colours turn bluer once the light curve starts to drop. An IR catastrophe is theoretically predicted to occur after several hundred days when the ejecta have cooled down to 1000--2000\,K \citep{axelrod1980a}, and has been confirmed at such a late epoch in SN~2011fe \citep{fransson2015a}. Given the spectroscopic evolution of SN~2002es, which continues to be similar to that of SN~1991bg, such a drastic change in the plasma state is strongly disfavoured. Finally, if the entire light-curve peak had merely been the result of shock breakout without any radioactive re-heating of the ejecta, a progenitor radius of $\sim15$\,R$_\odot$ would have been required. This is only feasible with an extended H envelope, but no H is observed in the spectra of SN~2002es. In summary, the rapid dimming of SN~2002es starting a month after maximum light remains a mystery. Other 02es-like SNe either do not have photometric coverage after +40\,d (SN~1999bh, PTF\,10ujn and PTF\,10acdh), or do not show a similar drop as SN~2002es (SN~2006bt, SN~2010lp, PTF\,10ops and iPTF\,14atg).

\subsection{Host galaxies and explosion environments}
\label{02es-like hosts}

Similar to their 91bg-like cousins, 02es-like events have a tendency to explode preferentially but not exclusively in massive, early-type host galaxies with $R$-band absolute magnitudes between $-19$ and $-22$ \citep{white2015a}. The explosion site of SN~2006bt was found at a large projected distance of 33.7\,kpc from its elliptical host galaxy \citep{foley2010a}. PTF\,10ops was even more extreme in this regard. \citet{maguire2011a} found no potential host galaxy in the surroundings of PTF\,10ops down to a limiting absolute magnitude of $-12$ in $R$, fainter than any known SN host. They concluded that PTF\,10ops was most likely associated with a massive galaxy at a projected distance of 148\,kpc, making this one of the most remote SNe discovered to date. Such extreme offsets are reminiscent of Ca-rich gap transients (Section~\ref{Ca-rich}), and impose strong constraints on the longevity of the progenitor systems.

\subsection{Explosion models}
\label{02es-like models}

Given the similarity between 02es-like and 91bg-like SNe, much of the model discussion from Section~\ref{91bg-like SNe} applies here as well. In particular, pure-deflagration models are disfavoured because of the strong mixing in the ejecta, which is in conflict with the stratified ejecta required to produce 91bg-like spectra \citep{hachinger2009a}. The main differences in 02es-like SNe are their somewhat larger \Nifs\ masses, meaning that the faintest \MCh\ delayed-detonation models are not necessarily excluded, and their broader light curves and partially lower velocities, which point at higher opacities and possibly a larger ejecta mass than in 91bg-like SNe.

Important constraints on progenitor systems and explosion models are provided by the early-time UV spike in iPTF\,14atg. In principle, there are several ways to produce a short-lived early UV flash in SN explosions, including the presence of radioactive material close to the surface of the ejecta at negligible optical depth \citep{diehl2014a}, the breakout of the SN shock from the progenitor star \citep{piro2010a,rabinak2012a,nakar2012a}, and an interaction of the SN ejecta with circumstellar material \citep{raskin2013a} or a companion star \citep{kasen2010a}. \citet{cao2015a} discussed most of these scenarios in the context of iPTF\,14atg, and concluded that except for ejecta\,--\,companion interaction all other models could be ruled out, suggesting that the progenitor of iPTF\,14atg must have had a non-degenerate companion. However, as pointed out by \citet{kromer2016a}, with respect to ejecta\,--\,CSM interaction, Cao et al. have only excluded a very specific CSM configuration, and neglect the possibilities of a more compact CSM with non-negligible optical depth in the shocked region \citep{raskin2013a} or a non-spherical CSM distribution.

A completely different approach to constrain the nature of 02es-like SNe concentrates on their spectroscopic similarity to 91bg-like SNe. \citet{maguire2011a} and \citet{ganeshalingam2012a} noted that a violent-merger model of two 0.9\,\Msun\ WDs by \citet{pakmor2010a} might be an attractive scenario for PTF\,10ops and SN~2002es. This model, originally meant to explain 91bg-like SNe, has too slowly-declining light curves to provide a convincing fit to 91bg-like SNe, but for 02es-like SNe this property is beneficial. \citet{kromer2013b,kromer2016a} presented revised versions of the model, i.e. mergers of a 0.9\,\Msun\ primary and a 0.76\,\Msun\ secondary at different metallicities. The slightly asymmetric mass configuration of these models leads to a higher \Nifs\ production, and both the light curves and spectra provide excellent matches to those of the 02es-like SNe~2010lp and iPTF\,14atg. Having unburned oxygen near the centre of the ejecta, these models might also potentially reproduce the [\OI] $\lambda\lambda6300,6364$ emission observed in the nebular spectra of SN~2010lp \citep{taubenberger2013b} and iPTF\,14atg \citep{kromer2016a}.

\citet{pakmor2013a} finally suggested that both 91bg-like and 02es-like SNe might stem from violent mergers whose primary CO WDs have a thin He layer on the surface. This He layer could be dynamically ignited during the merger process and trigger a subsequent C detonation in the core. This scenario should work with He as well as CO WD companions. He WD companions might result in 91bg-like SNe, while the significantly more massive CO WD companions would lead to a higher total ejecta mass and might explain 02es-like SNe.

\newpage

\section{Ca-rich gap transients \index{Ca-rich gap transients}}
\label{Ca-rich}

Traditionally, the classification of SNe was exclusively based on spectroscopic properties near maximum light. However, over the years it has unsurprisingly turned out that maximum-light spectra are not sufficient to capture and categorise the full diversity of phenomena revealed by more and more advanced observational campaigns. As a consequence, additional classification criteria based on light curves or the late-time evolution were introduced, and attempts were made to group together objects that were believed to share the same progenitors and explosion mechanisms. The group of Ca-rich gap transients constitutes an extreme example of this approach, as the assignment to this group is not at all based on the early-time spectroscopic properties (which can be quite diverse), but mostly on `Ca-rich' late-time spectra and a luminosity in the `gap' between novae and ordinary supernovae \citep{kasliwal2012a}. 

Ca-rich gap transients are one of the youngest SN classes. The first observations of such objects were carried out in the early 2000s \citep{filippenko2003a}, but it was only years later that Ca-rich gap transients were fully characterised as a separate class of objects \citep{perets2010a,kasliwal2012a}. Meanwhile, the `gap' is no longer empty, and encompasses SN~2005E \citep{perets2010a,kasliwal2012a}, SN~2005cz \citep{kawabata2010a,perets2011a}, PTF\,09dav \citep{sullivan2011a,kasliwal2012a}, SN~2007ke, SN~2010et\,/\,PTF\,10iuv and PTF\,11bij \citep{kasliwal2012a}, SN~2012hn \citep{valenti2014a}, and a few other Ca-rich transients reported by \citet{perets2010a}.

\begin{figure}[p]
\includegraphics[scale=1.0]{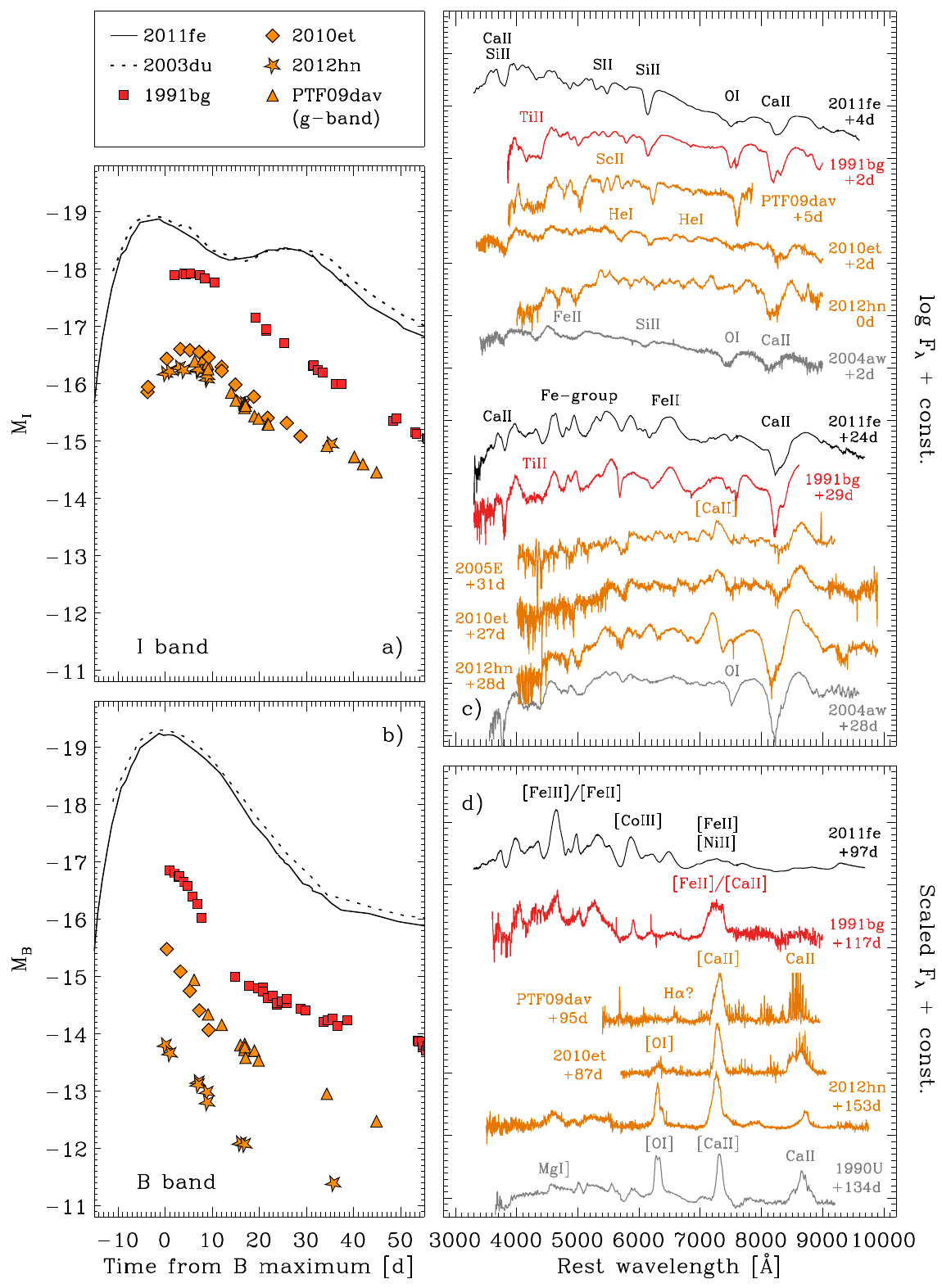}
\caption{Ca-rich gap transients \index{Ca-rich gap transients} in comparison to the normal SNe~Ia 2003du \citep{stanishev2007b} and 2011fe \citep{munari2013a,pereira2013a}, the subluminous SN~Ia 1991bg \citep[][see also Section~\ref{91bg-like SNe}]{filippenko1992b,leibundgut1993a,turatto1996a}, and the stripped-envelope core-collapse SNe~1990U and 2004aw \citep{taubenberger2006a}. a)~$I$-band light curves. b)~$B$-band light curves ($g$-band for PTF\,09dav). c)~Photospheric spectra. d)~Nebular spectra. References for individual Ca-rich gap transients are provided in the main text.}
\label{fig:5}
\end{figure}

\subsection{Light curves and peak luminosity}
\label{Ca-rich light curves}

Classical novae have $R$-band absolute peak magnitudes between $-5$ and $-10$, classical SNe are (with the exception of a few low-luminosity SNe~IIP) brighter than $-16$ at peak. The `gap' between these two luminosity regimes has until recently been relatively devoid of transients, but the advent of new high-cadence wide-field surveys such as PTF, PanSTARRS or OGLE has populated this strip from two sides, with luminous novae and subluminous SNe. Ca-rich gap transients are characterised by peak magnitudes between $-14.0$ and $-16.5$ \citep{perets2010a,kasliwal2012a}, and thus fall on the bright end of the gap. The morphology of their light curves is similar to 91bg-like SNe (Section~\ref{91bg light curves}) or fast-evolving SNe~Ib/c, with single peaks in the NIR bands, rise times of 9--15\,d, fast post-maximum decline and relatively red colours [$(B-V)_\mathrm{max} \sim 0.6$\,mag] (Fig.~\ref{fig:5}a,b).

\citet{waldman2011a} compared the bolometric light curve of SN~2005E (with a peak luminosity log($L_\mathrm{max}$)\,=\,41.5) to those of models featuring different compositions of radioactive species. It seems that a satisfactory fit can be obtained with a light curve powered exclusively by the \Nifs\ decay chain, at least if the total ejecta mass is adjusted accordingly (which Waldman et al. did not do). In that case, a \Nifs\ mass just below 0.01\,\Msun\ would be required. Alternatively, the light curve of SN~2005E could also be reproduced by a combination of mostly $^{48}$Cr, $^{52}$Fe and $^{44}$Ti decay, in which case, however, a ridiculously high $^{44}$Ti mass in excess of 1\,\Msun\ would be needed \citep{waldman2011a,kasliwal2012a}. Without a more precise knowledge of the ejecta mass, this degeneracy cannot be resolved easily. Finally, since the light curves of most Ca-rich gap transients look quite similar, the results obtained for SN~2005E are also applicable to other members of this class.

\subsection{Spectroscopic characteristics}
\label{Ca-rich spectra}

Since the group definition of Ca-rich gap transients is not based on early-time spectroscopic properties, it is not surprising to encounter some diversity among the group members. For most of them, however, a common scheme emerges. These objects are characterised by He-rich and H-free early-time spectra, formally leading to an SN~Ib classification \citep{perets2010a,kasliwal2012a}. Indeed, their photospheric spectra bear strong similarity to those of stripped core-collapse SNe of Type Ib/c \citep[][see Fig.~\ref{fig:5}c]{kawabata2010a,perets2010a,kasliwal2012a,valenti2014a}, showing lines of \HeI, \OI, \MgII, \CaII\ and maybe \SiII, \TiII\ and \FeII, with photospheric velocities of 10\,000--11\,000\,\kms\ at maximum light. Ca-rich gap transients turn nebular very early, so that after 50--100\,d the ejecta seem to be transparent. The spectra are then dominated by strong [\CaII] $\lambda7291,7324$ emission (Fig.~\ref{fig:5}d). The \CaII\ NIR triplet and [\OI] $\lambda\lambda6300,6364$ are usually also detected but comparatively weak, and [Fe] emission lines are entirely absent. With these characteristics, Ca-rich gap transients are different from both SNe~Ia and stripped core-collapse SNe during the nebular phase, but closer to core-collapse SNe (which show basically the same lines, just with stronger [\OI]) than to SNe~Ia with their Fe-dominated nebular spectra. The detection of [\OI] emission, in particular, is normally considered a hallmark feature of core-collapse explosions. In fact, the spectroscopic evolution led \citet{kawabata2010a} to interpret SN~2005cz as a core-collapse explosion of a low-mass ($\sim$10\,\Msun), envelope-stripped star, in spite of its location in an elliptical galaxy. This, however, was before the class of Ca-rich transients had been fully established.

Following \citet{arnett1982a}, an ejecta mass of 0.25--0.41\,\Msun\ and a kinetic energy of 4.4--$7.2 \times 10^{50}$\,erg have been estimated for SN~2005E on the basis of its early light curve and ejecta velocity \citep{perets2010a}. The same authors modelled a +62\,d spectrum of SN~2005E with a nebular spectrum synthesis code to obtain an estimate of the total ejecta mass of 0.275\,\Msun, half of which would be Ca. Such a low total mass and such a large Ca fraction (about two orders of magnitude higher than in any other SN) were unprecedented.  As a caveat it should be mentioned, however, that the +62\,d spectrum was not fully nebular, and a continuum contribution was subtracted before the modelling, introducing considerable uncertainties in the derived numbers. Moreover, for the Arnett formalism, an assumption had to be made regarding the mean ejecta opacity. Choosing this to be the same as in normal SNe~Ia led to the quoted numbers, while a lower mean opacity would result in higher ejecta-mass and kinetic-energy estimates.

That the distinction between Ca-rich transients and faint SNe~Ib/c is not always sharp is exemplified by SN~2012hn \citep{valenti2014a}. This SN had peak magnitudes similar to other Ca-rich transients, and was thus 1--2\,mag fainter than normal SNe~Ib/c. The flux below 5500\,\AA\ was strongly suppressed already at maximum light owing to \FeII, \TiII\ and \CrII\ line blanketing, and no \HeI\ lines were detected, so that the early-time spectra showed little resemblance to those of other Ca-rich gap transients (Fig~\ref{fig:5}c). However, [\CaII] emission started to emerge already at maximum light, and one month after peak [\CaII] and the \CaII\ NIR triplet were by far the strongest features in the spectrum. In a truly nebular spectrum taken 150\,d after maximum, on the other hand, [\OI] emission was almost equally prominent as [\CaII], and their ratio was not far from what is typically found in stripped core-collapse SNe (Fig.~\ref{fig:5}d). \citet{valenti2014a} estimated the ejecta mass to lie between $\sim$0.7 and $\sim$2.1\,\Msun, with the large range of values being due to the fairly unconstrained mean ejecta opacity.

Finally, PTF\,09dav was also very different from other Ca-rich transients (but also from SN~2012hn) at early phases. \citet{sullivan2011a} instead found similarities with 91bg-like SNe, but unusually low ejecta velocities and curious \ScII\ lines. A more in-depth discussion of the spectroscopic and photometric properties can be found in Section~\ref{PTF09dav}. Three months after maximum the spectrum of PTF\,09dav has changed completely and is entirely dominated by broad [\CaII] emission, like  those of more canonical Ca-rich gap transients \citep{kasliwal2012a}. However, again there are a few remarkable differences: [\OI] emission is not detected at all in PTF\,09dav, and the only feature not caused by Ca is a weak and relatively narrow \Ha\ line, whose width suggests an origin in a shocked H-rich CSM. With an estimated ejecta mass of 0.36\,\Msun\ \citep{sullivan2011a}, PTF\,09dav aligns well with the masses derived for other Ca-rich gap transients \citep{kasliwal2012a}.

\subsection{Host galaxies and explosion environments}
\label{Ca-rich hosts}

Up to this point, the reader may have wondered why a thermonuclear origin is favoured for Ca-rich gap transients by most authors, although the spectra and light curves bear stronger similarity to core-collapse SNe than to SNe~Ia. One reason are certainly the rather low inferred ejecta masses which seem to be at odds with massive-star progenitors. The main reason, however, are the explosion sites of many of these objects.

Ca-rich transients have a high fraction of early-type galaxies among their hosts: about 50\% of this still rather small sample of objects exploded in E or S0 galaxies \citep{perets2010a,kasliwal2012a}, to be compared to $\sim22$\% of SNe~Ia and at most $\sim1$\% of core-collapse SNe \citep{li2011a}. The stellar populations of these E and S0 hosts are predominantly old, and even if there had been recent star formation in a few of the hosts so that some young, massive stars are present, they are highly unlikely to be the progenitors of Ca-rich gap transients: if Ca-rich gap transients had massive progenitors, they should be much more frequently observed in actively star-forming galaxies, where those stars are by orders of magnitude more abundant.

The argument becomes even stronger when taking into account the locations of Ca-rich transients within their hosts. The distribution of explosion sites seems to be strongly skewed towards large galactocentric distances \citep{perets2010a,kasliwal2012a,yuan2013a,lyman2014a}. Projected distances between 20 and 40\,kpc have been derived for several objects, and distances above 5\,kpc are quite common. \citet{yuan2013a} have shown that for half of the Ca-rich transients essentially 100\% of the host-galaxy $K$-band light is enclosed by isophotes passing through their locations.

There are two basic scenarios of how stars can end up at such remote locations: either they were formed in situ, or they were expelled from their hosts.\newline
\textit{i) In-situ formation.} Deep pre- and post-explosion images and spectra have been investigated for a number of Ca-rich gap transients at large projected distances. None of them shows any concentration of light (satellite galaxy, globular cluster, massive star, star-forming region) at the site of the explosion, to limiting absolute magnitudes between $-5.3$ and $-12.4$ \citep{vandyk2003a,perets2010a,perets2011a,kasliwal2012a,lyman2014a}. The deepest of these limits effectively rule out the presence of globular clusters (GCs) and all except for the very faintest known dwarf galaxies \citep{lyman2014a}. In fact, \citet{yuan2013a} have shown that the radial distribution of GCs would be in good agreement with that of Ca-rich gap transients, but yet they do not favour this scenario given the available detection limits. Any association of massive stars \citep{lyman2014a} or ongoing star formation at the locations of Ca-rich transients can be excluded on safe grounds. Hence, if the progenitors of Ca-rich transients are really formed in situ, they must belong to a long-lived population of rather low-mass stars. \citet{yuan2013a} have suggested that the locations of Ca-rich transients could be consistent with relatively metal-poor, low-mass Halo stars, but \citet{lyman2014a} doubt any association with a genuine Halo population.\newline
\textit{ii) Runaway systems.} The alternative to in-situ formation are progenitor systems formed in more central regions of the hosts galaxies, and then kicked out by gravitational interaction with a supermassive black hole \citep[SMBH;][]{foley2015a} or as a consequence of a preceding supernova explosion \citep{lyman2014a}. Massive stars can be excluded as progenitors even in this case, since for realistic kick velocities of a few hundred \kms\ the progenitors' life times have to be close to 100\,Myr to reach the locations of the most remote Ca-rich transients observed. Less massive single stars, on the other hand, do not explode, so binarity is probably a key property. \citet{lyman2014a} have pointed out strong parallels between Ca-rich transients and short-duration $\gamma$-ray bursts (SGRBs) in terms of host-galaxy morphologies and offsets from their hosts. These SGRBs are usually thought to be mergers of two neutron stars or a neutron star (NS) and a black hole, which received their kicks by the two preceding core-collapse explosions within the system.

\subsection{Explosion models}
\label{Ca-rich models}

Explosion scenarios for Ca-rich gap transients need to be compatible with the main characteristics of these objects as inferred from observations: a low mass of \Nifs, a probably rather low ejecta mass, a rapid spectroscopic and photometric evolution, a Ca-rich chemical composition (though the exact degree of enhancement is not yet robustly established), the presence of He, and their locations at sometimes very large distances from the centres of their hosts.

The spectra, which in most cases resemble SNe~Ib at early phases and show strong [\CaII] next to weak [\OI] lines at nebular epochs, might suggest a core-collapse origin of Ca-rich transients \citep{kawabata2010a}, but as explained in Section~\ref{Ca-rich hosts} this scenario is disfavoured by their locations far from other young, massive stars or star-forming regions. In the realm of thermonuclear explosions, on the other hand, it is very unlikely from a nucleosynthesis perspective that Ca-rich transients are deflagrations or detonations of CO WDs like normal SNe~Ia. Neither the abundance of He and Ca, nor the paucity of Fe, Co and Ni could be explained with that scenario. Instead, more exotic models have to be invoked.

One scenario frequently discussed in the context of Ca-rich transients is the .Ia model \citep{bildsten2007a,shen2010a,waldman2011a,woosley2011b,sim2012a}. Similar to the double-detonation model for normal SNe~Ia, a detonation is ignited in an accreted or natively present He layer on the surface of a CO or ONe WD. The difference is that in the .Ia model the emergent shock wave fails to ignite the core. The mass of the He layer, and hence the ejecta mass, can range from a few times 0.01 to $\sim$0.3\,\Msun\ \citep{shen2010a,waldman2011a}, in fair agreement with the numbers estimated for Ca-rich gap transients, though a bit on the low side. The nucleosynthesis in the He detonation mostly proceeds to Ca, Ti and Cr, with only a relatively small amount of lighter IMEs and Fe\,/\,Co\,/\,Ni produced, again in agreement with the inferred ejecta composition of Ca-rich transients. Light curves and spectra of .Ia models vary strongly with the initial setup, and are sensitive to both the mass of the He shell and that of the core of the WD (which affects the density of the He layer). An acceptable match to the bolometric light curves of Ca-rich transients can probably be obtained if these parameters are chosen accordingly. For instance, \citet{waldman2011a} found that the detonation of a 0.2\,\Msun\ He shell on top of a 0.45\,\Msun\ WD provides a good match to the bolometric peak of SN~2005E, but a somewhat too rapid decline. For the same initial conditions, \citet{sim2012a} obtained a slightly more complete nucleosynthesis, favouring the production of the short-lived $^{48}$Cr at the expense of $^{44}$Ti, resulting in a significantly brighter light-curve peak. 

\citet{sim2012a} have also shown that in sufficiently low-mass WDs the ignition of a second detonation (either in the centre of the CO core via converging shocks or on the interface between the core and the He layer) does not enhance the production of \Nifs\ or other radioactive species by much, but leads to a significantly larger ejecta mass. The effect are explosions which are slightly more luminous at peak and have (at least for centrally ignited second detonations) broader light curves. This possibility thus increases the parameter space for models to explain faint and (more or less) rapidly declining transients.

Another way to produce faint and rapidly-declining transients is an accretion-induced collapse (AIC) of a WD. \citet{darbha2010a} have shown that such events can eject about the right \Nifs\ mass to reproduce the peak luminosity of Ca-rich gap transients, but that the total ejecta mass is very low, so that the ejecta velocities are too high, the light curve declines too rapidly and the ejecta composition does not match that inferred for Ca-rich transients. \citet{darbha2010a} speculated that a better match might be achieved by enshrouded AICs, where the Ni-dominated ejecta collide with surrounding unburned CO-rich material. The impact may trigger partial burning of C and O to IMEs, and the increased mass and opacity would decelerate the ejecta and slow down the light-curve evolution.

A completely different approach tries to constrain possible explosion scenarios for Ca-rich gap transients from their unusual locations. \citet{lyman2014a} proposed the progenitors of Ca-rich transients to most likely be runaway stars in binary systems. One of the constituents of the system would be a NS, which received a strong kick upon formation in a core-collapse SN. The other constituent would possibly be a WD. Based on calculations by \citet{metzger2012a}, NS--WD mergers would produce faint optical counterparts, whose peak luminosity and energetics would be in rough agreement with Ca-rich transients. However, the predicted Ca abundance is much lower than what \citet{perets2010a} inferred for SN~2005E, which is a potential shortcoming of this scenario. Moreover, it appears questionable if a binary system would remain bound in a core-collapse SN explosion where the newly formed NS receives a kick of several hundred \kms, required to reach the explosion sites of some Ca-rich transients tens of kpc away from the centres of their hosts.

\citet{foley2015a} instead suggested the kicks to be caused by gravitational interaction with central SMBHs. They found that many host galaxies of Ca-rich transients show signs of recent mergers, so that they might host binary SMBHs, which should be effective in ejecting stellar systems. If a binary system of a CO and a He WD that would normally not merge within a Hubble time is not only ejected, but also hardened in the course of the ejection process, a CO-He WD merger might take place tens to hundreds of Myr later at a large galactocentric distance. While this scenario could nicely explain the unusual locations of Ca-rich transients, it remains to be shown that CO-He WD mergers would result in explosions with the same low luminosity, low inferred ejecta masses and unusual nucleosynthesis pattern as revealed by Ca-rich gap transients.

\newpage

\section{Are the fastest decliners \index{Fast-declining transients} thermonuclear SNe?}
\label{02bj-like}

The objects that are the subject of this section are extremely few in number, and still show considerable diversity. Forcing them into one class would be pointless, and so this section is rather a collection of weird individuals that share a steep rise before and a tremendously rapid and deep decline after peak, and for which a thermonuclear origin seems to be at least an option.

\begin{figure}[p]
\includegraphics[scale=1.0]{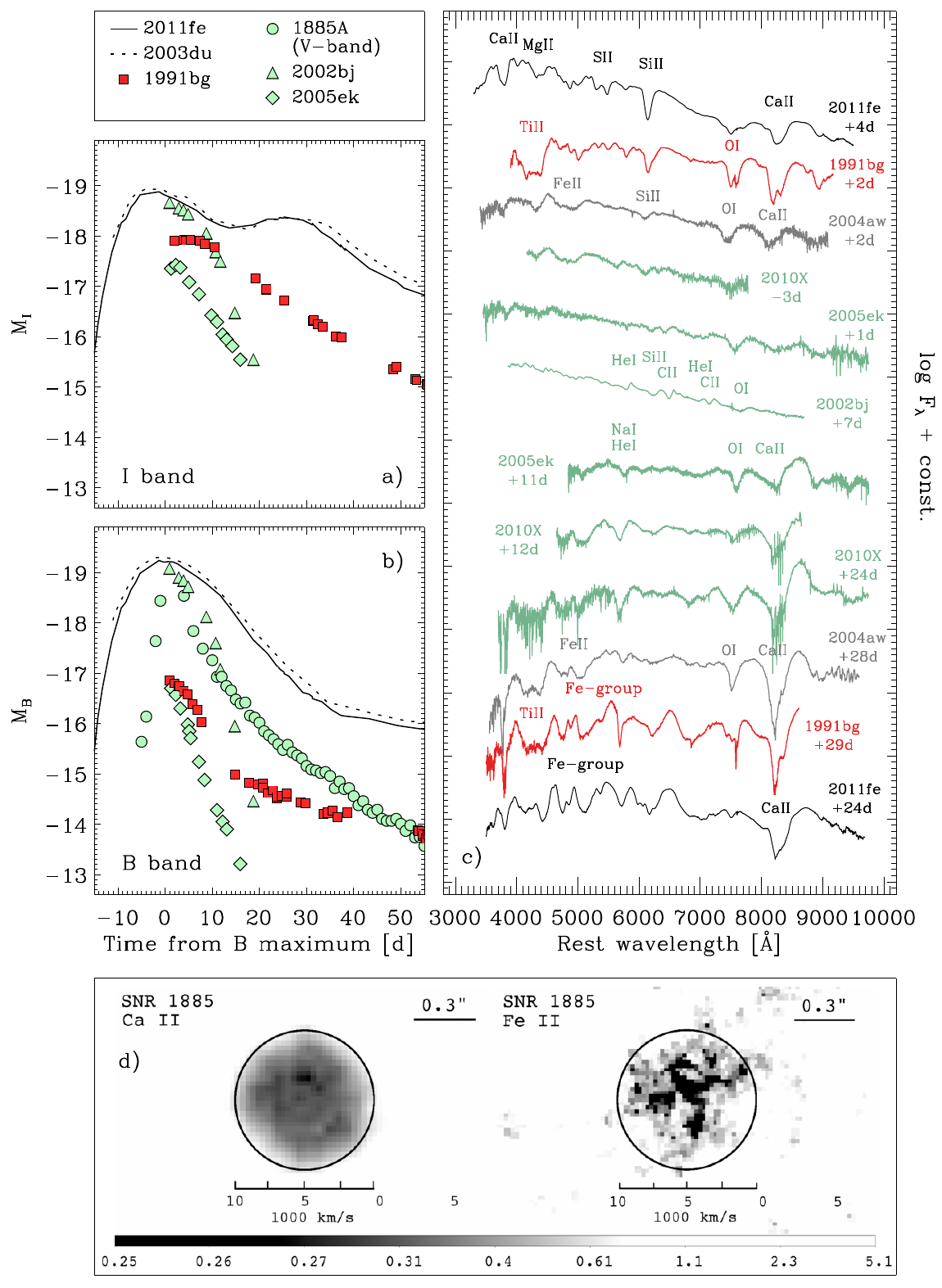}
\caption{Fast-declining transients \index{Fast-declining transients} in comparison to the normal SNe~Ia 2003du \citep{stanishev2007b} and 2011fe \citep{munari2013a,pereira2013a}, the subluminous SN~Ia 1991bg \citep[][see also Section~\ref{91bg-like SNe}]{filippenko1992b,leibundgut1993a,turatto1996a}, and the stripped-envelope core-collapse SN~2004aw \citep{taubenberger2006a}. a)~$I$-band light curves. b)~$B$-band light curves ($V$-band for SN~1885A). c)~Photospheric spectra. d)~The remnant of SN~1885A, seen in absorption against the bulge of M31 in near-UV lines of \CaII\ and \FeII. \copyright AAS. Reproduced with permission from \citet{fesen2015a}. References for individual fast-declining transients are provided in the main text.}
\label{fig:6}
\end{figure}

\subsection{SN~1885A -- observations in the 19th century}
\label{85A}

SN~1885A in M31, also known as S Andromedae, was the closest and brightest SN visible from the northern hemisphere for centuries. Discovered independently on 1885 August 19 and 20 by Isaac Ward and Ernst Hartwig, respectively, and thereafter followed by many astronomers, it reached 6th magnitude at peak \citep[see][for references to all historical reports]{devaucouleurs1985a}. Had it been in an isolated position, it would have been a naked-eye target, but since it exploded at a projected distance of only 16\,arcsec from the centre of M31 \citep{fesen1989a}, a small telescope or binoculars were needed to spot it. Lacking modern detector technology and even sufficiently fast photographic emulsions, the observations were performed visually, estimating the brightness and colour of the transient relative to those of other well-known stars. De Vaucouleurs and Corwin (1985) did an admirable job in carefully collecting, assessing and homogenising over 500 of these 100-year-old observations and converting them into a modern $V$-band light curve (Fig.~\ref{fig:6}b), $B-V$ colour curve and spectroscopic information.

At the time of discovery, SN~1885A was only about a day before peak, and several non-detections before August 17 allow to constrain the rise time of the transient to be extremely short, a few days at most. At the distance of M31 and corrected for Galactic extinction, an observed peak magnitude of $\sim5.9$ corresponds to an absolute $V$-band magnitude of about $-18.7$, so the transient was \textit{not} faint at peak. After maximum light, SN~1885A declined by $\sim3.5$\,mag within the first month [$\Delta m_{15}(V) \approx 2.3$\,mag; \citealt{devaucouleurs1985a,vandenbergh2002a}] before settling on a more gentle slope reminiscent of the \Cofs\ tail seen in other SNe. However, the transition between peak and tail is much more gradual than usual, casting some doubts on the correctness of the interpretation as a radioactivity-powered tail. The colour of SN~1885A was always red, but especially at early phases. A colour curve reconstructed by \citet{devaucouleurs1985a} shows a $B-V$ colour of $\sim1.3$\,mag at peak, which remains constant for $\sim20$\,d, and then turns bluer, reaching $B-V\sim0.9$\,mag at $+30$\,d. Such a colour evolution is unmatched by any other SN observed ever since. Verbal descriptions of spectroscopic properties and their evolution \citep{devaucouleurs1985a} agree on the presence of a (quasi-)continuum with superimposed broad lines, which is typical for SNe. Balmer lines were not detected, but apart from that an unambiguous line identification on this basis seems difficult.

\subsection{The remnant of SN~1885A}
\label{85A remnant}

The remnant of SN~1885A was found by \citet{fesen1989a} through ground-based narrow-band near-UV imaging. The remnant appears as a little dark spot superimposed on the bulge of M31, at a projected distance of only $\sim60$\,pc from the centre of the galaxy. To be seen in absorption, it has to be on the near side of M31, back-lit by the bulge. All previous searches for emission lines, starting already with Walter Baade in the 1940s \citep{osterbrock2001a}, had failed. Subsequent HST imaging and spectroscopy of the remnant \citep{fesen1999a,fesen2007a,fesen2015a,hamilton2000a} allowed a much more detailed study of its size, composition and geometry. The remnant is still in free expansion, and has reached a diameter of 0.7--0.8\,arcsec, implying an outer-edge ejecta velocity of $\sim$12\,500\,\kms\ \citep{fesen1999a}. The strongest absorption (down to $\sim20$\% transmission) is observed in the \CaII\ H\&K resonance lines, where the remnant looks nearly spherical on global scales \citep[][see Fig.~\ref{fig:6}d]{fesen1999a,fesen2007a,fesen2015a}. Other detected absorption lines are from Ca\,{\sc i}, \FeI\ and \FeII. In Ca\,{\sc i} and \FeI, which trace the regions with the highest Ca and Fe abundances, the remnant appears slightly smaller than in \CaII\ and lopsided, indicating a radial ionisation gradient in the ejecta and an off-centre peak in the Ca and Fe distribution \citep{fesen2007a}. \FeII, on the other hand, shows a number of plumes or filaments (Fig.~\ref{fig:6}d), and is quite different from the other ions in its spatial distribution \citep{fesen2015a}. 

The SN~1885A remnant has not been detected in X-rays. Adopting the ISM gas density in the bulge of M31 measured by \citet{li2009a} from diffuse X-ray emission, an \MCh\ SN~Ia remnant should be orders of magnitude brighter than the observed limits \citep{perets2011b}. Perets et al. therefore argued that either SN~1885A exploded in a local void, or the ejecta mass of the transient was much lower than that of a canonical \MCh\ SN~Ia.

\subsection{SN~1939B}
\label{39B}

Photometric data for SN~1939B have been compiled from historic sources by \citet{leibundgut1991b}. The light curve of this SN is similarly narrow as that of SN~1885A, with a somewhat slower rise \citep{devaucouleurs1985a} and an even slightly faster initial decline. \citet{perets2011b} measured a \dm15\ of 2.6. About two weeks after maximum, the light curve settles on a tail which is most likely \Cofs-powered. The transition between peak and tail phase is much sharper than in SN~1885A, and more similar to what is observed in other Type~I SNe. SN~1939B was probably even more luminous than SN~1885A, reaching $M_{B,\mathrm{max}} \lesssim -19$ (Fig.~\ref{fig:1}). In a $+22$\,d spectrum of SN~1939B published by \citet{perets2011b}, similarities with normal Type~Ia and Ib/c SNe are evident below 5500\,\AA, but an unusual behaviour is seen above that wavelength. This might, however, be attributed to flux-calibration problems of the photographic spectrum.

\subsection{SN~2002bj}
\label{02bj}

The rise time of SN~2002bj is constrained to be less than one week by a non-detection 7\,d before maximum down to a limiting magnitude 3.5\,mag below peak \citep{poznanski2010a}. However, not only the rise but also the decline is very rapid, and steepening further after $\sim10$\,d. Within 18\,d from maximum, SN~2002bj drops by 4.5\,mag in the $B$ band \citep[][see Fig.~\ref{fig:6}b]{poznanski2010a}, and \citet{perets2011b} measured a remarkable \dm15\ of 3.2. There is no indication of a flattening of the light curves or a settling on a radioactive tail out to 20\,d after maximum when the follow-up ends.
Like SNe~1885A and 1939B, SN~2002bj features a rather high peak luminosity $\log L_{BVRI} \approx 43.0$ ($M_{B,\,\mathrm{max}} \approx -19$; \citealt{poznanski2010a}). Over 15--20\,d after maximum light, there is only a mild colour evolution, and the $B-V$ colour remains comparatively blue at $\sim0$\,mag.

Spectra of SN~2002bj have been obtained at 7 and 11\,d after maximum \citep{poznanski2010a}. They show a blue continuum with numerous weak P-Cygni lines superimposed (Fig.~\ref{fig:6}c). Modelling the spectra, \citet{poznanski2010a} identified carbon, oxygen, IMEs and helium. \CII\ lines are very prominent and the \HeI\ detection is also unambiguous. \CaII\ lines, in contrast, are very weak. No evidence was found for IGEs or hydrogen. With P-Cygni absorptions blueshifted by 4000\,\kms\ the ejecta velocities are extraordinarily low. Finally, a polarisation spectrum showed no line polarisation exceeding 0.2\%, indicating that composition asymmetries are small.

Applying Arnett's rule \citep{arnett1982a} and scaling relations to SN~2002bj, \citet{poznanski2010a} found numbers that were difficult to reconcile. Their estimate of the \Nifs\ mass of 0.15--0.25\,\Msun\ equalled or exceeded that of the total ejecta mass of $\sim0.15$\,\Msun. Even if for the lower-limit \Nifs\ mass of 0.15\,\Msun\ a direct contradiction could possibly be avoided, it would imply that the ejecta consist to 100\% of \Nifs, which is inconsistent with the observed spectra which show IMEs but lack any evidence of IGEs. Moreover, for the estimated \Nifs\ mass a much brighter light-curve tail would be expected. \citet{poznanski2010a} therefore speculated that an energy source other than \Nifs\ decay is required to power the light curve of SN~2002bj, and suggested a contribution by $^{48}$Cr decay as a possible solution.

\subsection{SNe~2005ek and 2010X}
\label{10X}

With SNe~2005ek \citep{drout2013a} and 2010X \citep{kasliwal2010a} there are finally two SNe with sufficiently similar behaviour to be discussed together.

For SN~2010X no multi-band photometry is available around maximum light. The existing $r$-band light curve, however, is morphologically nearly identical to that of SN~2002bj. SN~2005ek also has very similar light curves, but with a slightly more linear decline right after peak (Fig.~\ref{fig:6}a,b), and with extensive multi-band coverage during the first 15--20\,d after maximum which allowed \citet{drout2013a} to measure a \dm15\ of 3.5.  The rise times of SNe~2005ek and 2010X are not as tightly constrained as that of SN~2002bj, and could well be longer than 10\,d. Like SN~2002bj, SNe~2005ek and 2010X show no flattening of their decline out to $+17$\,d. For SN~2005ek, however, two late-time detections at $+38$ and $+68$\,d prove the presence of a radioactive tail which likely starts around day 20. The optical colours of SN~2005ek are significantly redder than those of SN~2002bj, and show a stronger time evolution to the red (with $B-V$ rising from 0.2\,mag at peak to 0.9\,mag on day 15). What distinguishes SNe~2005ek and 2010X from the other SNe discussed in this section, however, is their significantly lower peak luminosity ($M_{R,\,\mathrm{max}} \approx -17.3$ and $-17.0$, respectively).

The spectroscopic evolution of SNe~2005ek ($-1$ to $+11$\,d; \citealt{drout2013a}) and 2010X ($-4$ to $+34$\,d; \citealt{kasliwal2010a}) appears similar to those of normal stripped-envelope core-collapse SNe of Type Ic (Fig.~\ref{fig:6}c), with lines of \OI, \CII, \CaII\, \NaI, \MgII, \SiII\, \TiII\ and \FeII\ detected through spectral modelling \citep{drout2013a,kleiser2014a}. The evolutionary time scales of the spectra are relatively fast, in line with the narrow light curves. In fact, \citet{drout2013a} noted slight flux excesses in a $+9$\,d spectrum of SN~2005ek with respect to a synthetic fit, which might be attributed to emerging lines of [\OI] $\lambda5577$, [\OI] $\lambda\lambda6300,6364$ and [\CaII] $\lambda\lambda7291,7324$. This would be one of the earliest detections of nebular emission in a SN of any type. In both SNe `normal' ejecta velocities of about 8000--10\,000\,\kms\ are measured a few days after maximum light, at least twice as high as in SN~2002bj. In general, despite the photometric similarities, spectroscopically SNe~2005ek and 2010X have little in common with SN~2002bj.

Estimates of $M_\mathrm{Ni}$, $M_\mathrm{ej}$ and $E_\mathrm{kin}$ for SNe~2005ek and 2010X yield a much more consistent picture of low-mass and low-energy \Nifs-powered explosions than in the case of SN~2002bj, mainly due to the lower luminosities. For SN~2010X, \citet{kasliwal2010a} analytically derived a \Nifs\ mass of 0.02\,\Msun, a total ejecta mass of 0.16\,\Msun, and a kinetic energy of $\sim1.5 \times 10^{50}$\,erg. Spectral modelling of SN~2005ek provided satisfactory results for an ejecta mass of 0.3\,\Msun\ and a kinetic energy of $2.5 \times 10^{50}$\,erg \citep{drout2013a}.

\subsection{Host galaxies and environments}
\label{02bj-like hosts}

While SNe~2002bj, 2005ek and 2010X exploded in spiral galaxies with presumably heterogeneous stellar populations \citep{poznanski2010a,kasliwal2010a}, SN~1939B was found in an elliptical galaxy and SN~1885A was associated with the bulge of M31 with a characteristic age of the stellar population of $\sim10$\,Gyr \citep{perets2011b}. For Perets et al. this high fraction of occurrence in old stellar populations favours WDs as the progenitors for these SNe. For SNe~1885A and 1939B this is certainly correct. However, considering the diversity among the four objects discussed in this section, it is unclear whether this conclusion applies to SNe~2002bj, 2005ek and 2010X.

\subsection{Explosion models}
\label{02bj-like models}

There are several approaches to constrain the explosion mechanism(s) underlying the group of rapidly declining Type I SNe discussed above. \citet{fesen2015a} tried to make use of the spatially resolved abundance structure of the SN~1885A remnant (Fig.~\ref{fig:6}d). Since the remnant is still expanding freely, the structure imprinted by the explosion is conserved, providing a spatially resolved image of the SN ejecta. From apparent plumes in the distribution of \FeII\ absorption, Fesen et al. infer the presence of a deflagration phase, and favour a delayed-detonation mechanism. However, the rapidly declining light curve of SN~1885A seems inconsistent with delayed-detonation explosions \citep[e.g.][]{seitenzahl2013a}. Moreover, \citet{perets2011b} argued that an \MCh\ explosion was probably difficult to reconcile with the observed X-ray limits for the remnant, supporting a sub-Chandrasekhar ejecta mass. 

Other attempts to identify suitable explosion models revolve around ways to produce low ejecta masses, since most analytical estimates and numerical modelling of light curves and spectra suggest ejecta masses between 0.15 and 0.30\,\Msun\ \citep{poznanski2010a,kasliwal2010a,drout2013a}. The first such model was invoked by \citet{chevalier1988a}, who found good agreement between a synthetic light curve for a He surface detonation of \citet{woosley1986b} and the observed light curve of SN~1885A. That model ejected 0.26\,\Msun\ of material, 0.19\,\Msun\ of which were \Nifs. Similarly, \citet{poznanski2010a} proposed a .Ia explosion model for SN~2002bj, i.e. a thermonuclear flash in a low-mass He shell on the surface of a WD \citep{bildsten2007a,shen2010a}. These flashes produce transients with peak absolute magnitudes between $-15$ and $-18$ and rise times $<10$\,d \citep{shen2010a}. Their rise is definitely fast enough to be compatible with SN~2002bj, but they are not quite luminous enough, and the decline of the light curves flattens out too early. The same conclusion was reached by \citet{waldman2011a}, whose series of He surface detonations cover an even wider range of outcomes (but primarily extend to lower peak luminosities). 

Another problematic aspect of .Ia SNe is their nucleosynthesis, as pointed out by \citet{drout2013a}. The ejecta of .Ia SNe consist mostly of He and elements between Ca and Ni, but show very low abundances of C, O and IMEs such as Mg, Si and S, which have all been detected in the spectra of SNe~2002bj, 2005ek and 2010X, and according to spectral modelling of SN~2005ek \citep{drout2013a} even dominate the ejecta by mass. Synthetic spectra for He detonations \citep{waldman2011a,sim2012a}, in contrast, are almost exclusively shaped by elements from Ca to Cr, and are very red even at early epochs owing to strong line blanketing at short wavelengths. \citet{drout2013a} therefore suggested that edge-lit double-detonations of low-mass WDs as presented by \citet{sim2012a} might provide a better match in terms of ejecta composition than pure He-shell detonations. However, as shown by \citet{sim2012a}, the spectroscopic difference between edge-lit double detonations and pure He-shell detonations is marginal at least for their models, since the early spectra are always formed in the ashes of the He shell.

\citet{drout2013a} compared the light curve of SN~2005ek also to other thermonuclear models with low ejecta mass. They concluded that accretion-induced collapse \citep[e.g.][]{darbha2010a} and NS--NS mergers \citep[e.g.][]{metzger2010a} could be ruled out based on their low luminosities and extremely narrow light curves, but that WD--NS or WD--BH mergers \citep[e.g.][]{metzger2012a} might be possible candidates, although also these models tend to be too faint.

However, also core-collapse models are being discussed as explanation for at least some of the objects in this section. Based on the spectroscopic similarity of SNe~2005ek and 2010X with SNe~Ic such as SN~1994I, \citet{drout2013a} proposed that the binary-interaction scenario for envelope stripping might also work for stars with an initial mass just slightly above the limit for iron core collapse. Indeed, \citet{tauris2015a} studied close binary systems of a He star and an accreting NS, and found that these can lead to ultra-stripped core-collapse SNe with ejecta masses of just a few tenths of a solar mass. The low optical depth of the ejecta would then result in strong $\gamma$-ray leakage and short retention times for optical photons, both of which would help to explain the rapid decline of the light curves.

An alternative core-collapse scenario has been proposed by \citet{kleiser2014a}, and this is the only model so far that seems capable of reproducing the observed narrow light curves without the need for a small ejecta mass. The idea is that of a rather normal stripped-envelope core-collapse explosion with very little radioactive material. The peak light curve would be powered by the recombination of oxygen and other abundant elements initially ionised by the SN shock, in analogy to H recombination in SNe~IIP. Once the ejecta have recombined and the stored energy is exhausted, the end of the oxygen-plateau phase is reached and the light curve drops rapidly. A small amount of radioactive material could then still produce a radioactive tail.

\newpage

\section{`Super-Chandrasekhar' SNe~Ia: \index{`Super-Chandrasekhar' SNe~Ia} the mass puzzle}
\label{super-Ch}

For a long time there was general consensus that CO WDs explode at the very latest when approaching a mass of $\sim$1.4\,\Msun, known as the Chandrasekhar-mass stability limit. The \MCh\ limit is very fundamental \citep{chandrasekhar1931a}, denoting the mass where the degeneracy pressure of the relativistic electron gas loses against the self-gravity of the WD, leading to a collapse. Even luminous 91T-like SNe (see Section~\ref{91T-like SNe}) have been shown to be consistent with \MCh\ progenitors \citep[e.g.][]{stritzinger2006a,scalzo2014a}. It was only during the past decade that a few objects were discovered whose enormous luminosities, broad light curves and moderately low ejecta velocities seemed to be in conflict with explosions of 1.4\,\Msun\ WDs and instead suggested larger ejecta masses. The explosions were therefore dubbed `super-Chandrasekhar' SNe~Ia.

As of now, the class of `super-Chandrasekhar' objects is still small, including SNe~2003fg \citep{howell2006a}, 2004gu \citep{contreras2010a}, 2006gz \citep{hicken2007a,maeda2009a}, 2007if \citep{scalzo2010a,yuan2010a,childress2011a,taubenberger2013a}, 2009dc \citep{yamanaka2009a,tanaka2010a,silverman2011a,taubenberger2011a,taubenberger2013a,hachinger2012a} and SN~2012dn \citep{chakradhari2014a,parrent2016a}. SNe~2006gz and 2012dn have many properties in common with other `super-Chandrasekhar' SNe~Ia, but challenge any group definition based on the sheer luminosity owing to their comparatively moderate photometric properties.

\begin{figure}[p]
\includegraphics[scale=1.0]{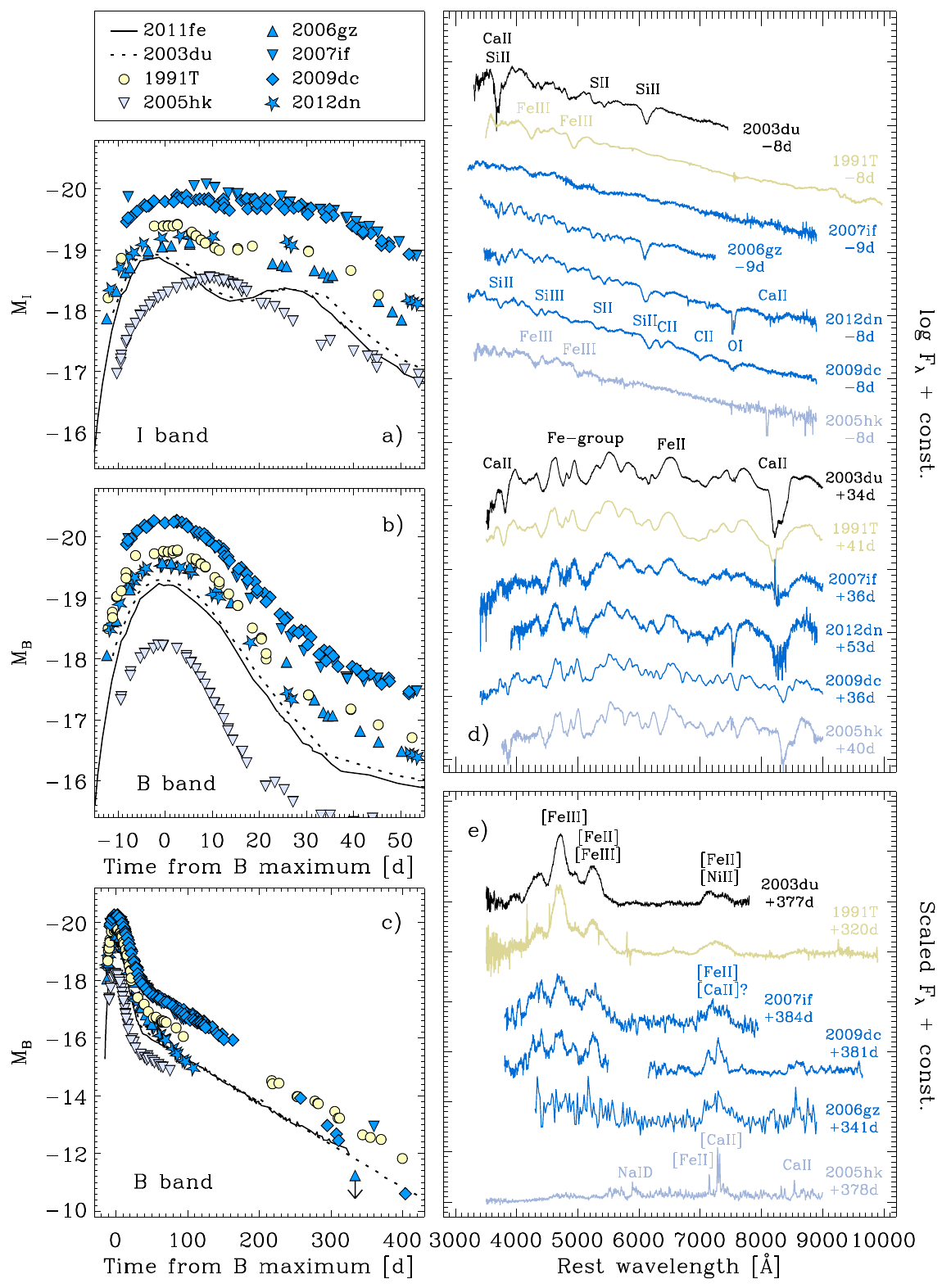}
\caption{`Super-Chandrasekhar' SNe~Ia \index{`Super-Chandrasekhar' SNe~Ia} in comparison to the normal SNe~Ia 2003du \citep{stanishev2007b} and 2011fe \citep{munari2013a,pereira2013a}, the SN~Iax 2005hk \citep{phillips2007a,silverman2012a} and SN~1991T \citep[][see also Section~\ref{91T-like SNe}]{mazzali1995a,gomez1998a,lira1998a,altavilla2004a,silverman2012a}. a)~$I$-band light curves at early times. b)~$B$-band light curves at early times. c)~Late-time $B$-band light curves. d)~Photospheric spectra. e)~Nebular spectra. References for individual `super-Chandrasekhar' SNe~Ia are provided in the main text.}
\label{fig:7}
\end{figure}

\subsection{Early-time light curves and peak luminosity}
\label{super-Ch light curves}

The overall light-curve morphology of `super-Chandrasekhar' SNe~Ia is quite similar to that of normal SNe~Ia, with a smooth, rounded peak followed by a radioactive-tail phase. The main difference lies in the peak luminosity and the time scales of the light-curve rise and decline, but also certain colours show a characteristic evolution distinguishing `super-Chandrasekhar' from ordinary SNe~Ia. $B$-band peak absolute magnitudes typically lie between $-19.5$ and $-20.4$ \citep[][see Figs.~\ref{fig:1} and \ref{fig:7}b]{hicken2007a,scalzo2010a}, quasi-bolometric peak luminosities $L_\mathrm{uvoir}$ between 43.2 and 43.5, outshining average normal SNe~Ia by a factor $\sim$2 \citep{taubenberger2011a}. The light curves evolve very slowly. For SN~2009dc, \citet{silverman2011a} reported a first detection on an image taken 21\,d before $B$-band maximum, and estimated a rise time of $23\pm2$\,d, significantly longer than in normal SNe~Ia (17--20\,d; \citealt{conley2006a,hayden2010a}). Also the decline rates after peak are among the slowest ever observed for SNe~Ia, with \dm15\ between 0.7 and 0.9. 

Applying Arnett's rule \citep{arnett1982a} to the quasi-bolometric light-curve peaks of `super-Chandrasekhar' SNe~Ia leads to high \Nifs-mass estimates, not only because of the high luminosity, but also because of the long rise time. For SN~2009dc, \citet{yamanaka2009a}, \citet{silverman2011a} and \citet{taubenberger2011a} find \Nifs\ masses of $1.8\pm0.5$\,\Msun, which by itself is at best marginally compatible with the Chandrasekhar mass of a non-rotating WD. If the ejecta are not entirely made up of \Nifs\ (which they are not), there is no way to reconcile the peak light curve of SN~2009dc with the explosion of a 1.4\,\Msun\ WD unless other energy sources help to power it. \citet{scalzo2010a} reached a similar conclusion for SN~2007if by fitting the peak and the tail of the light curve. They inferred a \Nifs\ mass larger than $\sim1.3$\,\Msun\ and a total ejecta mass in excess of $\sim1.8$\,\Msun\ at $3\sigma$ confidence level, with most likely numbers of $M_\mathrm{Ni}\approx1.6$\,\Msun\ and $M_\mathrm{ej}\approx2.2$\,\Msun.

Apart from high peak luminosities and broad light curves, `super-Chandrasekhar' SNe~Ia are characterised by an unusually high $U$-band and UV flux and blue UV\,$-$\,optical colours \citep{brown2014a}, at least compared to normal SNe~Ia where that region is strongly suppressed due to heavy line blanketing by IGEs. The $V-I$ and $V-\mathrm{NIR}$ colour evolution of `super-Chandrasekhar' SNe~Ia is also very different from normal SNe~Ia, which is related to the fact that these objects do not show distinct secondary maxima in the $IJHK$ bands, but rather broad, delayed, plateau-like peaks \citep[][see Fig.~\ref{fig:7}b]{taubenberger2011a,friedman2015a}.

Most other SNe~Ia, from luminous 91T-like objects to moderately subluminous 86G-like events, appear to be very good standard candles in the NIR \citep{wood-vasey2008a,krisciunas2009a,kattner2012a}. `Super-Chandrasekhar' SNe~Ia defy this trend, being overluminous by $\sim$1\,mag also in these bands \citep{taubenberger2011a}. Similarly, they are apparently not good standardisable candles at optical wavelengths \citep{howell2006a,silverman2011a,taubenberger2011a}, being too luminous for their light-curve shape and thus lying systematically `above' the Phillips relation \citep{phillips1993a,phillips1999a}.

All the characteristics described above apply without exception to SNe~2003fg, 2007if and 2009dc. SNe~2006gz and 2012dn share most of these properties, from a slow light-curve decline over a single-peaked, broad $I$-band light curve to unusually high UV luminosities \citep{hicken2007a,chakradhari2014a,brown2014a}. However, with peak absolute magnitudes $M_B \approx -19.5$ they are significantly fainter than other `super-Chandrasekhar' SNe~Ia, and in fact just marginally more luminous than normal SNe~Ia.

\subsection{Early-time spectroscopic properties}
\label{super-Ch spectra}

SNe~2003fg, 2006gz, 2009dc and 2012dn show pre-maximum spectra dominated by IMEs and unburned material, with lines of IGEs probably making some contribution below 5000\,\AA\ \citep[][see Fig.~\ref{fig:7}d]{hicken2007a,silverman2011a,chakradhari2014a,parrent2016a}. \CaII\ lines are weaker than in normal SNe~Ia, \CII\ and \OI\ more prominent. \CII\ $\lambda6580$ and $\lambda7234$, in particular, are unusually strong and persistent, being visible not only up to $\sim$10\,d before peak as in normal SNe~Ia, but even two weeks after maximum \citep{silverman2011a,taubenberger2011a,chakradhari2014a,parrent2016a}. SN~2007if differs significantly from the rest, with early spectra dominated by a blue continuum with only weak \FeIII\ and IME lines superimposed, suggesting a high ionisation state similar to 91T-like SNe (Section~\ref{91T-like SNe}). However, though weaker than in the other `super-Chandrasekhar' SNe~Ia, \CII\ lines are also detected in SN~2007if, and similarly persistent even past maximum light \citep{scalzo2010a,yuan2010a}. All `super-Chandrasekhar' SNe~Ia show a very blue SED and a high UV flux in early spectra, indicating little line blanketing by IGEs.

High-velocity features have not been found in `super-Chandrasekhar' SNe~Ia even at the earliest epochs \citep{taubenberger2011a}, while they are very common in normal SNe~Ia, at least in \CaII\ \citep{mazzali2005a}. But irrespective of the absence of high-velocity features, also the `photospheric' velocities are moderate to low, ranging from 8000 to 12\,000\,\kms\ for \SiII\ $\lambda6355$ and from 7500 to 10\,000\,\kms\ for \CII\ $\lambda6580$ at maximum light \citep{hicken2007a,taubenberger2011a,chakradhari2014a}. These velocities imply a low kinetic energy per mass. Coupled with the large energy release from synthesising  the amount of \Nifs\ inferred from the light-curve peak, this would necessitate a large ejecta mass, with a significant amount of unburned material left. For SN~2009dc, \citet{taubenberger2011a} estimated that a total ejecta mass close to 3\,\Msun\ would be required to keep the velocities as low as observed if really  1.5\,\Msun\ of \Nifs\ or more had been synthesised in the explosion.

About 2--4 weeks after maximum, as the photosphere recedes into the Fe core of the ejecta, the spectra of `super-Chandrasekhar' SNe~Ia become increasingly \FeII-dominated. The same transition takes place in all normal SNe~Ia as well, but typically a bit earlier. During the following weeks the spectra of the moderate-velocity specimens of `super-Chandrasekhar' SNe such as SNe~2007if and 2012dn are very similar to those of normal SNe~Ia, while those of the low-velocity SN~2009dc resemble closely those of SNe~Iax, with a wealth of resolved lines which are blended in normal SNe~Ia (Fig.~\ref{fig:7}d). 

For SN~2009dc a polarisation spectrum was recorded by \citet{tanaka2010a} a few days after maximum light. It showed no significant continuum polarisation ($<0.3$\%), indicating minimum deviations from spherical symmetry on global scales. Despite the lack of typically strongly polarised high-velocity features, a polarisation up to 0.7\% was observed in lines of \SiII\ and \CaII, which points at some inhomogeneities in the spatial distributions of those elements.

\subsection{Evolution during the nebular phase}
\label{super-Ch nebular}

Some (but not all) `super-Chandrasekhar' SNe~Ia exhibit an enhanced fading after a certain point in time in all optical bands (Fig.~\ref{fig:7}c). In SN~2009dc, which has the best light-curve coverage of all `super-Chandrasekhar' SNe~Ia, the rapid dimming starts $\sim$200\,d after maximum and lasts for at least 150\,d, during which an additional fading by $\sim$1.2\,mag is observed compared to the behaviour expected for a \Cofs-decay powered light curve with increasing $\gamma$-ray losses \citep{silverman2011a,taubenberger2011a}. For SN~2006gz the light-curve coverage is limited, but one year after maximum the quasi-bolometric light curve is suppressed by $\sim$2.5\,mag \citep{maeda2009a}, making the SN even significantly less luminous than ordinary SNe~Ia. SN~2012dn shows signs of an inflection point in the light curves as early as 70\,d after maximum \citep{chakradhari2014a}, but no real late-time photometry has been published to date. SN~2007if, on the other hand, shows no unexpected dimming, and its light curve remains on the usual \Cofs-decay tail out to one year after maximum \citep{taubenberger2013a}. 

The substantial dimming observed in SN~2006gz goes hand in hand with a strong suppression of the blue part of the spectrum \citep[][see Fig.~\ref{fig:7}e]{maeda2009a}. \citet{taubenberger2013a} showed that both effects, the light-curve drop and the suppression of blue flux, can be consistently explained by additional reddening in SNe~2006gz and 2009dc at late times, and speculated that dust formation -- though never seen in other thermonuclear SNe -- might have taken place in these events at different phases in a carbon-rich shell outside the Fe core. They estimated that a dust mass of few times $10^{-4}$\,\Msun\ would be sufficient to explain the dimming observed in SN~2009dc.

Before the putative dust formation sets in, the bolometric light-curve tail can be used to estimate \Nifs\ and ejecta masses. Although the luminosity difference to normal SNe~Ia during this phase is even larger than at peak, the \Nifs\ and ejecta masses derived from the tail are more modest. The reason is that the largest part of the increased luminosity during the tail phase does not come from a higher \Nifs\ mass, but from more efficient $\gamma$-ray trapping due to low ejecta velocities and hence high densities \citep{taubenberger2013a}. 

Spectra taken during the early nebular phase (100--200\,d after peak) are available for SNe~2007if and 2009dc \citep{silverman2011a,taubenberger2011a,blondin2012a} and show the same [\FeII], [\FeIII] and [\CoIII] emission features as those of normal SNe~Ia. However, as a consequence of the lower ejecta velocities, lines that are normally blended are partially resolved in SN~2009dc. One year after maximum, the spectra of `super-Chandrasekhar' SNe~Ia are still dominated by forbidden Fe lines, but the hallmark [\FeIII] feature at 4700\,\AA\  is very weak \citep[][see Fig.~\ref{fig:7}e]{silverman2011a,taubenberger2013a}. In normal SNe~Ia the flux ratio between the [\FeIII]-dominated 4700\,\AA\ feature and the [\FeII]-dominated 5200\,\AA\ feature lies between 1.3 and 1.8, in SNe~2007if and 2009dc between 1.0 and 1.1 \citep{taubenberger2013a}. This points at a low ionisation state, probably related to high ejecta densities.

\subsection{Host galaxies and explosion environments}
\label{super-Ch hosts}

`Super-Chandrasekhar' SNe show a tendency to explode in rather low-mass galaxies \citep{taubenberger2011a}, and those events that were found in more massive galaxies typically exploded in relatively remote locations. Taken together, this indicates that the metallicity at all explosion sites was probably low, and that low metallicity may in fact be a key ingredient in the progenitor evolution \citep{khan2011a}. SN~2007if, in particular, went off in an extreme dwarf galaxy ($M_g = -14.45$) with the lowest oxygen abundance [$12+\log(\mathrm{O/H}) = 8.01$] of any spectroscopically measured SN~Ia host galaxy \citep{childress2011a}. Hence, there might not only be a Malmquist, but also a metallicity bias favouring `super-Chandrasekhar' SNe in high-$z$ SN~Ia samples, constituting a potential source of systematic error in SN~Ia cosmology.

\subsection{Explosion models}
\label{super-Ch models}

The initial idea to explain `super-Chandrasekhar' SNe, promoted by \citet{howell2006a}, was that of a rapidly spinning WD whose limiting mass could quite significantly exceed 1.4\,\Msun\ owing to the stabilising action of centrifugal forces \citep{langer2000a,yoon2005b}. When a $\sim$2\,\Msun\ WD explodes, so the theory went, it could produce the required 1.2--1.5\,\Msun\ of \Nifs. \citet{hachisu2012a} even suggested, based on more refined binary evolution scenarios, that a heavily rotating WD could grow to 2.3--2.7\,\Msun\ before exploding, setting the stage even for the most extreme `super-Chandrasekhar' SNe~2007if and 2009dc. The presence of a strongly quantising magnetic field also may \citep{das2013b} or may not \citep{chamel2013a} be able to stabilise WDs up to such high masses. Finally, \citet{hicken2007a} proposed that a WD merger would be an easier way for a WD to gain such a high mass than accretion in a single-degenerate scenario.

However, as shown by the explosion simulations for a rapidly rotating 2\,\Msun\ WD by \citet{pfannes2010a,pfannes2010b}, this model is probably not suited to explain `super-Chandrasekhar' SNe~Ia. It might well produce superluminous transients, but fails in terms of explosion energetics and nucleosynthesis. If the flame proceeds subsonically \citep{pfannes2010a}, the differential rotation inhibits a flame propagation in the equatorial direction, and only 0.5\,\Msun\ of IGEs are produced in total. Moreover, most of the ejecta mass below 3000\,\kms\ is made up by unburned C and O, which is in clear conflict with the Fe-dominated nebular spectra of `super-Chandrasekhar' SNe. If the flame instead propagates supersonically, the differential rotation is less of an obstacle, and a very powerful explosion emerges. in this `AWD3det' model of \citet{pfannes2010b} 1.5 out of 2.0\,\Msun\ of the progenitor are burned into \Nifs, and only 0.08\,\Msun\ in an equatorial torus remain unburned, most of it ending up above 12\,000\,\kms. Spectrum-synthesis calculations by \citet{hachinger2012a} have subsequently shown that the AWD3det model produces too high ejecta velocities and has way too much burned material at high velocity for being a good match with SN~2009dc which is characterised by low velocities. 

The failure of the AWD3det model clearly shows that if the luminosity of SN~2009dc is to be generated by the decay of $\gtrsim1.5$\,\Msun\ of \Nifs, a total mass of $\sim3$\,\Msun\ is required to keep the ejecta velocities sufficiently low, and most of the additional mass would have to remain unburned \citep{hachinger2012a}. However, \citet{taubenberger2013a} have shown that the 3\,\Msun\ model of \citet{hachinger2012a} would be inconsistent with the bolometric light curve of SN~2009dc during the radioactive tail (before dust formation occurs). The high mass and low ejecta velocities would lead to very efficient $\gamma$-ray trapping, making the model light curve too bright by almost a factor 2 between 100 and 200\,d. From this consideration, it actually appears that there is no \Nifs\ mass -- ejecta mass combination that can consistently explain the peak luminosity, tail luminosity and ejecta velocities observed in SN~2009dc, at least without tremendous fine-tuning of the density profile or multi-dimensional effects.

A multi-dimensional effect that has been discussed quite early in the context of `super-Chandrasekhar' SNe~Ia is an explosion with an off-centre Ni distribution \citep{hillebrandt2007a}. These models exhibit lines of sight along which the peak luminosity is boosted, and others along which it is depressed. However, along the boosted lines of sight the peak is attained earlier and the post-maximum light curve declines faster than seen from other directions, in disagreement with observations of `super-Chandrasekhar' SNe~Ia. Also, there would be no effect on the light-curve tail, when the ejecta become optically thin, while in observed `super-Chandrasekhar' SNe the luminosity during the tail is even more enhanced compared to normal SNe~Ia than at peak.

A way out of this dilemma would be a scenario where the light curve is not entirely Ni-powered. The possibility of an enshrouded explosion along the lines of the `tamped detonations' of \citet{khokhlov1993b}, where the ejecta are decelerated by interaction with a surrounding circumstellar medium, was already discussed by \citet{hicken2007a}, \citet{scalzo2010a} and \citet{taubenberger2011a}. \citet{hachinger2012a} took this a step further by allowing for a contribution to the luminosity of SN~2009dc in the form of an underlying blackbody-like continuum in their spectral models, which led to superior fitting results. A potential problem with all interaction-aided models is that there are no clear interaction signatures in the light curves (sudden kinks or drops) and spectra (narrow or intermediate-width emission lines) of any `super-Chandrasekhar' SN~Ia. This could be achieved if the interaction phase is already over at the time of the first observations. There is no hydrogen observed in the spectra, so the surrounding material would have to be H-free. This excludes thermonuclear explosions of AGB-star cores \citep{taubenberger2011a} or explosions of WDs within common envelopes, leaving mergers of CO WDs with a rather compact CO-rich CSM as a viable scenario \citep{hicken2007a,taubenberger2011a,taubenberger2013a}. 

Motivated by the need to reproduce the tail light curve of SN~2009dc (see above), \citet{taubenberger2013a} proposed an explosion of a 1.4\,\Msun\ WD enshrouded by $\sim$0.6--0.7\,\Msun\ of CO material (the debris of the former companion WD), with a \Nifs\ mass of $\sim$1.0\,\Msun. In this model, the ejecta velocities and the tail luminosity would match those of SN~2009dc, but the radioactively powered light curve alone falls short of reproducing the light-curve peak in optical bands. To that end, a significant boost would have to come from X-ray and UV photons emitted by the shock soon after explosion, which would have to be trapped, reprocessed to longer wavelengths and released on photon-diffusion time scales. Whether this mechanism can actually work, remains to be tested with detailed simulations. However, \citet{taubenberger2013a} pointed out that several other peculiarities of `super-Chandrasekhar' SNe might also be consistently explained within this scenario: the large amount of swept-up carbon could give rise to the strong and persistent \CII\ features, the high early-time UV luminosity may be the `afterglow' of the shock emission, and the high density and carbon content in the shocked CSM region may provide favourable conditions for dust formation once these layers have cooled down sufficiently.

\section*{Acknowledgements}

The author acknowledges support by project TRR\,33 `The Dark Universe' of the German Research Foundation (DFG), and thanks Markus Kromer and Suhail Dhawan for helpful discussions.

%\section*{Cross-References}
%
%\begin{itemize}
%\item Observational Classification of Supernovae
%\item Hydrogen-Poor Core Collapse Supernovae
%\item Type Ia supernovae
%\item Type Iax supernovae
%\item Supernovae from low and Intermediate Mass stars (0.8-8 Msun)
%\item Light Curves of Type I Supernovae
%\item Spectra of supernovae during the photospheric phase
%\item Nebular spectra of supernovae
%\item Interacting Supernovae and the Influence on Spectra and Light Curves
%\item Unusual supernovae and alternative power sources
%\item Introduction to Supernova Polarimetry
%\item Dynamical Mergers
%\item Violent Mergers
%\item Chandrasekhar mass explosions
%\item Nucleosynthesis in thermonuclear supernovae
%\item The Peak Luminosity-Decline Rate Relationship for Type Ia Supernovae
%\item The Infrared Hubble Diagram of Type Ia Supernovae
%
%\end{itemize}

\def\aj{AJ}%
          % Astronomical Journal
\def\araa{ARA\&A}%
          % Annual Review of Astron and Astrophys
\def\apj{ApJ}%
          % Astrophysical Journal
\def\apjl{ApJ}%
          % Astrophysical Journal, Letters
\def\apjs{ApJS}%
          % Astrophysical Journal, Supplement
\def\ao{Appl.~Opt.}%
          % Applied Optics
\def\apss{Ap\&SS}%
          % Astrophysics and Space Science
\def\aap{A\&A}%
          % Astronomy and Astrophysics
\def\aapr{A\&A~Rev.}%
          % Astronomy and Astrophysics Reviews
\def\aaps{A\&AS}%
          % Astronomy and Astrophysics, Supplement
\def\azh{AZh}%
          % Astronomicheskii Zhurnal
\def\baas{BAAS}%
          % Bulletin of the AAS
\def\jrasc{JRASC}%
          % Journal of the RAS of Canada
\def\memras{MmRAS}%
          % Memoirs of the RAS
\def\mnras{MNRAS}%
          % Monthly Notices of the RAS
\def\pra{Phys.~Rev.~A}%
          % Physical Review A: General Physics
\def\prb{Phys.~Rev.~B}%
          % Physical Review B: Solid State
\def\prc{Phys.~Rev.~C}%
          % Physical Review C
\def\prd{Phys.~Rev.~D}%
          % Physical Review D
\def\pre{Phys.~Rev.~E}%
          % Physical Review E
\def\prl{Phys.~Rev.~Lett.}%
          % Physical Review Letters
\def\pasp{PASP}%
          % Publications of the ASP
\def\pasj{PASJ}%
          % Publications of the ASJ
\def\qjras{QJRAS}%
          % Quarterly Journal of the RAS
\def\skytel{S\&T}%
          % Sky and Telescope
\def\solphys{Sol.~Phys.}%
          % Solar Physics
\def\sovast{Soviet~Ast.}%
          % Soviet Astronomy
\def\ssr{Space~Sci.~Rev.}%
          % Space Science Reviews
\def\zap{ZAp}%
          % Zeitschrift fuer Astrophysik
\def\nat{Nature}%
          % Nature
\def\iaucirc{IAU~Circ.}%
          % IAU Cirulars
\def\aplett{Astrophys.~Lett.}%
          % Astrophysics Letters
\def\apspr{Astrophys.~Space~Phys.~Res.}%
          % Astrophysics Space Physics Research
\def\bain{Bull.~Astron.~Inst.~Netherlands}%
          % Bulletin Astronomical Institute of the Netherlands
\def\fcp{Fund.~Cosmic~Phys.}%
          % Fundamental Cosmic Physics
\def\gca{Geochim.~Cosmochim.~Acta}%
          % Geochimica Cosmochimica Acta
\def\grl{Geophys.~Res.~Lett.}%
          % Geophysics Research Letters
\def\jcp{J.~Chem.~Phys.}%
          % Journal of Chemical Physics
\def\jgr{J.~Geophys.~Res.}%
          % Journal of Geophysics Research
\def\jqsrt{J.~Quant.~Spec.~Radiat.~Transf.}%
          % Journal of Quantitiative Spectroscopy and Radiative Trasfer
\def\memsai{Mem.~Soc.~Astron.~Italiana}%
          % Mem. Societa Astronomica Italiana
\def\nphysa{Nucl.~Phys.~A}%
          % Nuclear Physics A
\def\physrep{Phys.~Rep.}%
          % Physics Reports
\def\physscr{Phys.~Scr}%
          % Physica Scripta
\def\planss{Planet.~Space~Sci.}%
          % Planetary Space Science
\def\procspie{Proc.~SPIE}%
          % Proceedings of the SPIE
\let\astap=\aap
\let\apjlett=\apjl
\let\apjsupp=\apjs
\let\applopt=\ao

\footnotesize{
  \bibliographystyle{apj}

}

\end{document}